\documentclass[fleqn,usenatbib,]{mnras}
\usepackage{lineno}
\usepackage{newtxtext,newtxmath}

\usepackage[T1]{fontenc}
\usepackage{ae,aecompl}

\usepackage{graphicx}	
\usepackage{amsmath}	
\usepackage{amssymb}	
\usepackage{booktabs}   
\usepackage{longtable}  
\usepackage{rotating}
\usepackage{appendix}
\usepackage{float}
\usepackage[dvipsnames]{xcolor}
\usepackage{subcaption}
\usepackage{multicol}
\usepackage{wrapfig}
\PassOptionsToPackage{hyphens}{url}\usepackage{hyperref}
\usepackage{threeparttable}
\usepackage{booktabs}

\usepackage{etoolbox}
\makeatletter
\patchcmd\@combinedblfloats{\box\@outputbox}{\unvbox\@outputbox}{}{%
  \errmessage{\noexpand\@combinedblfloats could not be patched}%
}%
\makeatother
\usepackage{eso-pic}

\defcitealias{Smith2020}{S20}
\defcitealias{Kelly2012}{KK12}

\title[Supernova hosts in DES]{Supernova Host Galaxies in the Dark Energy Survey: \\ I. Deep Coadds, Photometry, and Stellar Masses}
\author[DES Collaboration]{
\parbox{\textwidth}{
\Large
P.~Wiseman,$^{1}$
M.~Smith,$^{1}$
M.~Childress,$^{1}$
L.~Kelsey,$^{1}$
A.~M\"oller,$^{2}$
R.~R.~Gupta,$^{3}$
E.~Swann,$^{4}$
C.~R.~Angus,$^{1,5}$
D.~Brout,$^{6,7}$
T.~M.~Davis,$^{8}$
R.~J.~Foley,$^{9}$
C.~Frohmaier,$^{4}$
L.~Galbany,$^{10}$
C.~P.~Guti\'errez,$^{1}$
R.~Kessler,$^{11,12}$
G.~F.~Lewis,$^{13}$
C.~Lidman,$^{14}$
E.~Macaulay,$^{4}$
R.~C.~Nichol,$^{4}$
M.~Pursiainen,$^{1}$
M.~Sako,$^{7}$
D.~Scolnic,$^{15}$
N.~E.~Sommer,$^{14}$
M.~Sullivan,$^{1}$
B.~E.~Tucker,$^{14}$
T.~M.~C.~Abbott,$^{16}$
M.~Aguena,$^{17,18}$
S.~Allam,$^{19}$
S.~Avila,$^{20}$
E.~Bertin,$^{21,22}$
D.~Brooks,$^{23}$
E.~Buckley-Geer,$^{19}$
D.~L.~Burke,$^{24,25}$
A.~Carnero~Rosell,$^{26,18}$
M.~Carrasco~Kind,$^{27,28}$
L.~N.~da Costa,$^{18,29}$
J.~De~Vicente,$^{26}$
S.~Desai,$^{30}$
H.~T.~Diehl,$^{19}$
P.~Doel,$^{23}$
T.~F.~Eifler,$^{31,32}$
S.~Everett,$^{9}$
P.~Fosalba,$^{33,34}$
J.~Frieman,$^{19,12}$
J.~Garc\'ia-Bellido,$^{20}$
E.~Gaztanaga,$^{33,34}$
D.~W.~Gerdes,$^{35,36}$
R.~A.~Gruendl,$^{27,28}$
J.~Gschwend,$^{18,29}$
W.~G.~Hartley,$^{23,37}$
S.~R.~Hinton,$^{8}$
D.~L.~Hollowood,$^{9}$
D.~J.~James,$^{38}$
K.~Kuehn,$^{39,40}$
N.~Kuropatkin,$^{19}$
M.~Lima,$^{17,18}$
M.~A.~G.~Maia,$^{18,29}$
M.~March,$^{7}$
P.~Martini,$^{41,42}$
P.~Melchior,$^{43}$
F.~Menanteau,$^{27,28}$
R.~Miquel,$^{44,45}$
R.~L.~C.~Ogando,$^{18,29}$
F.~Paz-Chinch\'{o}n,$^{27,28}$
A.~A.~Plazas,$^{43}$
A.~K.~Romer,$^{46}$
A.~Roodman,$^{24,25}$
E.~Sanchez,$^{26}$
V.~Scarpine,$^{19}$
S.~Serrano,$^{33,34}$
E.~Suchyta,$^{47}$
M.~E.~C.~Swanson,$^{28}$
G.~Tarle,$^{36}$
D.~Thomas,$^{4}$
D.~L.~Tucker,$^{19}$
T.~N.~Varga,$^{48,49}$
A.~R.~Walker,$^{16}$
and R.~Wilkinson$^{46}$
\begin{center} (DES Collaboration) \end{center}
}
\vspace{0.4cm}
\\
\parbox{\textwidth}{Affiliations are listed at the end of the paper}
}
\date{Accepted XXX. Received YYY; in original form ZZZ}

\pubyear{2019}

\begin{document}
\label{firstpage}
\pagerange{\pageref{firstpage}--\pageref{lastpage}}
\maketitle

\begin{abstract}
The five-year Dark Energy Survey supernova programme (DES-SN) is one of the largest and deepest transient surveys to date in terms of volume and number of supernovae. Identifying and characterising the host galaxies of transients plays a key role in their classification, the study of their formation mechanisms, and the cosmological analyses. To derive accurate host galaxy properties, we create depth-optimised coadds using single-epoch DES-SN images that are selected based on sky and atmospheric conditions. For each of the five DES-SN seasons, a separate coadd is made from the other 4 seasons such that each SN has a corresponding deep coadd with no contaminating SN emission. The coadds reach limiting magnitudes of order $\sim 27$ in $g$-band, and have a much smaller magnitude uncertainty than the previous DES-SN host templates, particularly for faint objects. We present the resulting multi-band photometry of host galaxies for samples of spectroscopically confirmed type Ia (SNe Ia), core-collapse (CCSNe), and superluminous (SLSNe) as well as rapidly evolving transients (RETs) discovered by DES-SN. We derive host galaxy stellar masses and probabilistically compare stellar-mass distributions  to samples from other surveys. We find that the DES spectroscopically confirmed sample of SNe Ia selects preferentially fewer high mass hosts at high redshift compared to other surveys, while at low redshift the distributions are consistent. DES CCSNe and SLSNe hosts are similar to other samples, while RET hosts are unlike the hosts of any other transients, although these differences have not been disentangled from selection effects.

\end{abstract}




\section{Introduction \label{sec:intro}}
The accelerating expansion of the Universe, hypothesised to be driven by an unknown dark energy, is one of the largest unsolved problems in physics, astronomy, and cosmology. The discoverers used type Ia supernovae (SNe Ia) as standardisable candles to measure distances across the cosmos \citep{Riess1998,Perlmutter1999}. Since then, the scale of sky surveys dedicated to improving upon the accuracy and precision of cosmological measurements has increased dramatically \citep[e.g.][]{Astier2006,Kessler2009,Conley2011,Suzuki2012,Betoule2014}. The Pantheon analysis \citep{Scolnic2018} included a sample of $>1000$ SNe Ia, and when combined with the cosmic microwave background (CMB) constraints from \citet{Planck2016} measured the dark energy equation-of-state parameter $w$ to a precision of $\sim 0.04$. The Dark Energy Survey Supernova programme (DES-SN) is in the process of building an even larger sample and is aiming to further reduce systematic uncertainties. The results from the first three years of the survey (DES3YR) have recently demonstrated the state-of-the-art precision capabilities of DES-SN (\citealt{DESCollaboration2018a}). The DES3YR analysis included a photometric pipeline to determine lightcurves of 207 SNe Ia \citep{Brout2019}, spectroscopy using a range of large telescopes \citep{DAndrea2018}, a comprehensive analysis of the systematic uncertainties \citep{Brout2019a}, a suite of simulations \citep{Kessler2019}, inclusion of chromatic corrections to the calibration \citep{Lasker2019}, and a measurement of the Hubble constant, $H_0$ \citep{Macaulay2019}.

SNe Ia cosmology has traditionally been performed with `spectroscopic samples', in which all SNe in the sample have been confirmed as SNe Ia by analysing a spectrum of the SN. As transient surveys probe larger areas with deeper observations, however, it not feasible to  classify all of the SNe spectroscopically. We thus define samples by classifying SNe `photometrically', principally using the lightcurve shape and colour to distinguish SNe Ia from core-collapse events using classifiers such as \texttt{pSNid} \citep{Sako2008}, \texttt{SuperNNova} \citep{Moller2019}, and \texttt{RAPID} \citep{Muthukrishna2019}.

In both spectroscopic and photometric samples, determination of the host galaxy associated with each SN is crucial. Firstly, narrow emission and/or absorption lines in the spectrum of a host galaxy provide a much more precise measurement of the redshift than the broader lines of the SN spectrum, allowing for smaller uncertainties on the redshift axis of the Hubble diagram. Redshifts from the hosts are improve the photometric classification of transients \citep[e.g.][]{Olmstead2014,Sako2014}, with classification accuracy of the \texttt{SuperNNova} classifier improving from 97\% to $>99\%$ with the addition of redshift \citep{Moller2019}. 
Secondly, even after brightness corrections are applied using known correlations in their lightcurve shape (stretch) and colour, a residual intrinsic scatter in their absolute peak brightness is still measured. There exist further correlations, between the properties of the SN host galaxy and the colour-and-stretch corrected brightness (or Hubble residual) of the SN \citep[e.g.][]{Sullivan2006,Rigault2013,Roman2018}. Of these, stellar mass is the most robust and easily measured, leading to the so-called `mass step'  correction \citep[e.g.][]{Kelly2010,Lampeitl2010a,Sullivan2010, Conley2011, Childress2014}. Understanding the driver behind, and correcting for, the mass step is the focus of significant ongoing work (\citealt{Roman2018,Rigault2018,Jones2018,Rose2019,Smith2020,Kelsey2020}), all of which requires accurate and precise galaxy photometry.

Host galaxy properties are important not only for cosmological measurements, but also in the quest to understand the SN explosions themselves. Most commonly used due to their observational ease and simplicity, particularly at higher redshift, are global host galaxy properties. These include stellar mass, age, and star-formation rate, and are derived from observations of the galaxy as a whole. For nearby, spatially resolved galaxies, especially those for which integral field spectroscopy (IFS) observations are available, local properties can provide an extra channel from which to inform the host study \citep[e.g.][]{Thoene2014,Kruehler2017,Galbany2018,Schady2019}. Local properties are typically analogues of the global properties, but are derived from a region smaller than the entire host galaxy, and are used to provide a more accurate representation of the properties of the particular stellar population from which the progenitor was born \citep[e.g.][]{Rigault2013,Roman2018}.

Galaxy properties are commonly used to infer the nature of transients. Events linked to massive stars tend to occur in star-forming galaxies, thermonuclear transients and compact object mergers occur more universally \citep[e.g.][]{Childress2013,Palmese2017}, and tidal disruption events (TDEs) often occur in post-starburst E+A  galaxies \citep{Arcavi2014,French2016,Kruehler2018}. More specifically, the myriad subclasses of SNe each show a preference toward certain host properties: among those associated with massive stars, the most energetic such as gamma-ray bursts (GRBs; e.g. \citealt{Fruchter2006, Perley2016b, Graham2017}), superluminous supernovae (SLSNe; e.g. \citealt{Neill2011,Angus2016,Chen2017}), and relativistic broadline SNe (Ic-bl; \citealt{Japelj2018,Modjaz2019}) typically occur in environments low in metallicity and stellar mass, and/or high in specific star-formation rate (star-formation rate (SFR) per unit stellar mass), while more typical core collapse SNe (CCSNe) are more agnostic \citep[e.g.][]{Anderson2010}. The relatively small numbers of objects in some of these samples mean selection effects are also at play and must be correctly accounted for when drawing conclusions about progenitor populations.

Host galaxy properties can be estimated from photometry, slit spectroscopy, and more recently IFS. While spectroscopy is able to provide more detailed information about the physical processes at play in the galaxies, it is expensive and time consuming. The magnitude limits of spectroscopy are relatively shallow, which is a limitation when dealing with SNe at high redshifts or in faint host galaxies. On the other hand, the nature of wide-field, untargeted searches such as DES means that there is by design a wealth of imaging of the host for each and every transient detected in the survey in the form of the single-epoch exposures. In order to detect transients, a template image is subtracted from each single epoch exposure in a technique known as difference imaging. During the DES science verification (SV; see \citealt{Jarvis2015,Rykoff2016,Bonnett2016} for a detailed description of the SV data\footnote{\href{http://des.ncsa.illinois.edu/releases/sva1}{des.ncsa.illinois.edu/releases/sva1}}), templates for difference imaging \citep{Kessler2015} were constructed from roughly three nights of observing. While the difference imaging templates were updated throughout the survey with data from each season, the original SV templates were used to determine host galaxy properties for spectroscopic target selection \citep{DAndrea2018} and in the cosmology analysis \citep{Brout2019a}. In this work, we improve upon those templates by building coadds from the full survey.

The main DES-SN survey consisted of five annual, six-month observing seasons with repeated, roughly seven-day cadence observations in each of ten pointings of the 2.7 deg$^2$ field-of-view Dark Energy Camera (DECam; \citealt{Flaugher2015}), denoted the SN fields (Section \ref{subsec:stacks}). With a total of $\sim$ 120 visits to each field by the end of the survey \citep{Diehl2016,Diehl2018}, it is possible to improve upon the SV templates by stacking single-epoch images into coadds. Such a method has been used in other repeat-observation surveys such as SDSS Stripe 82 \citep{Annis2014}. In building a deep host galaxy template for each SN, it is necessary to omit the epochs in which the SN is active. Typically this is done by building separate multi-season coadds, omitting each season in turn (e.g. Pan-STARRS; \citealt{Rest2014,Scolnic2018}). SNe fade by several orders of magnitude on the timescale of a year - SN2003hv was around 7 magnitudes fainter than at peak in all optical bands 300 days post-peak \citep{Leloudas2009}, while the equivalent decline for SN2012fr to 150 days was 5 magnitudes \citep{Contreras2018}. Thus for `normal' SNe Ia occurring at the end of a season, their contribution to the host galaxy flux in the subsequent season beginning $\sim6$ months later is negligible. For SLSNe, whose lightcurve durations often exceed that of a DES observing season, it can be necessary to exclude data from the subsequent season and as such these coadds may not be suitable for analysing the hosts of some of the SLSNe in DES-SN.

In this paper, we lay the foundations for the analysis of the host galaxies of the full DES-SN data set. We build a suite of depth-optimised coadds and perform diagnostic tests comparing these coadds to other catalogues, which is described in Section \ref{sec:data}. In Section \ref{sec:seds}, we analyse the host galaxies of various transients in DES-SN, focusing on fitting their spectral energy distributions (SEDs) with stellar population templates. In Section \ref{sec:results} we describe the results of the SED fitting and report host masses for various subsets of transients. We summarise with a discussion and conclusion, in Sections \ref{sec:discussion} and \ref{sec:conclusion} respectively. Throughout this paper we adopt a spatially-flat $\Lambda\mathrm{CDM}$ cosmological model with a matter density $\Omega_{\mathrm{m}} = 0.3$ and Hubble constant $H_0 = 70$ kms$^{-1}$ Mpc$^{-1}$. We use AB magnitudes \citep{Oke1983} and report uncertainties at the $1\sigma$ level unless otherwise stated.

\begin{figure}
\includegraphics[width=0.5\textwidth]{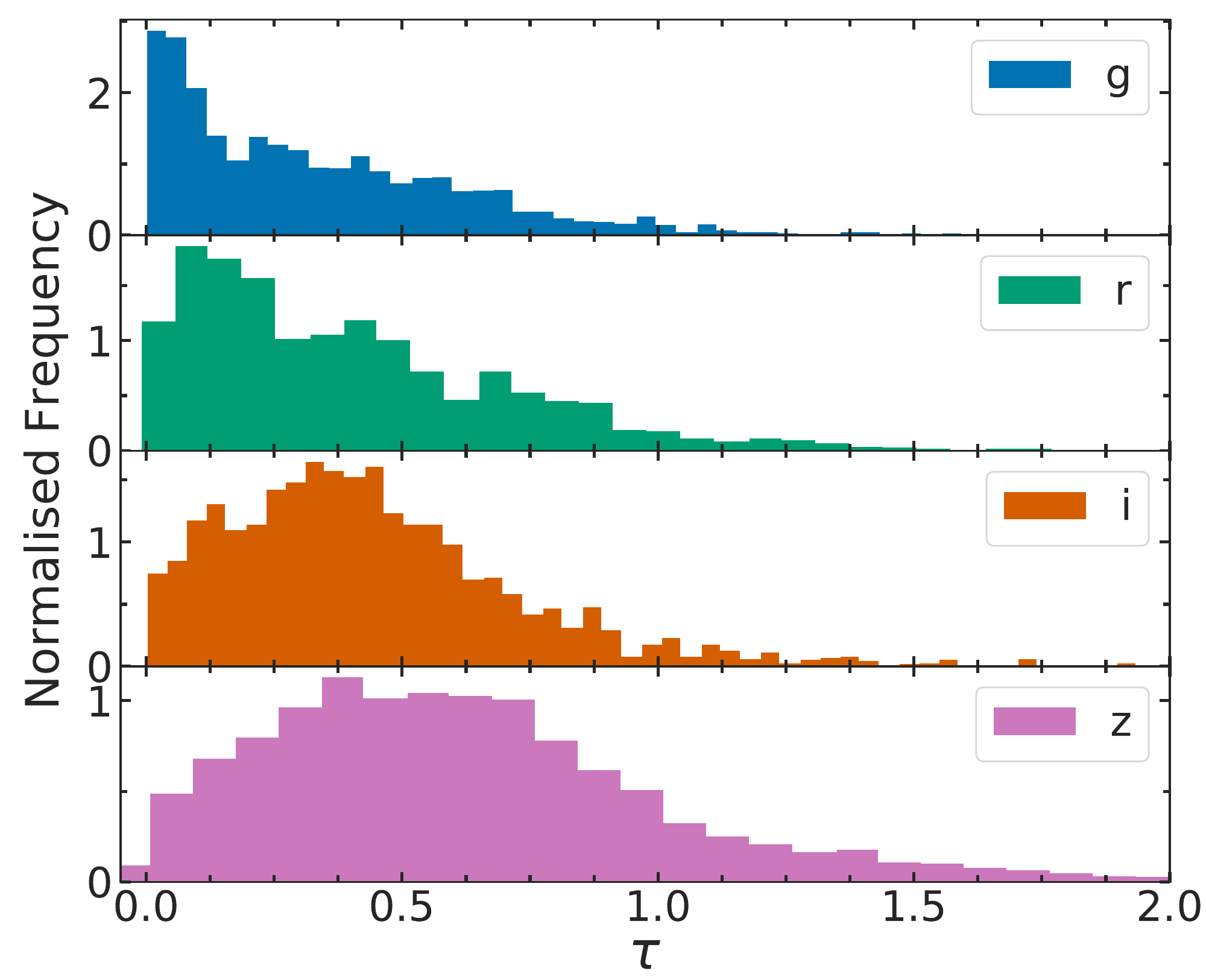}
\caption{Histograms for the distribution of effective exposure time ratio $\tau$ (the ratio between the true exposure time, and the effective exposure time based on the atmospheric and sky background conditions; Eq. \ref{eq:teff})  across all DES-SN exposures. The distributions peak at progressively higher values at higher wavelengths, meaning that a larger fraction of the $i$ and $z$ band exposures are closer to the fiducial `good' conditions.
\label{fig:teff_hist}}
\end{figure}

\begin{figure}
\includegraphics[width=0.5\textwidth]{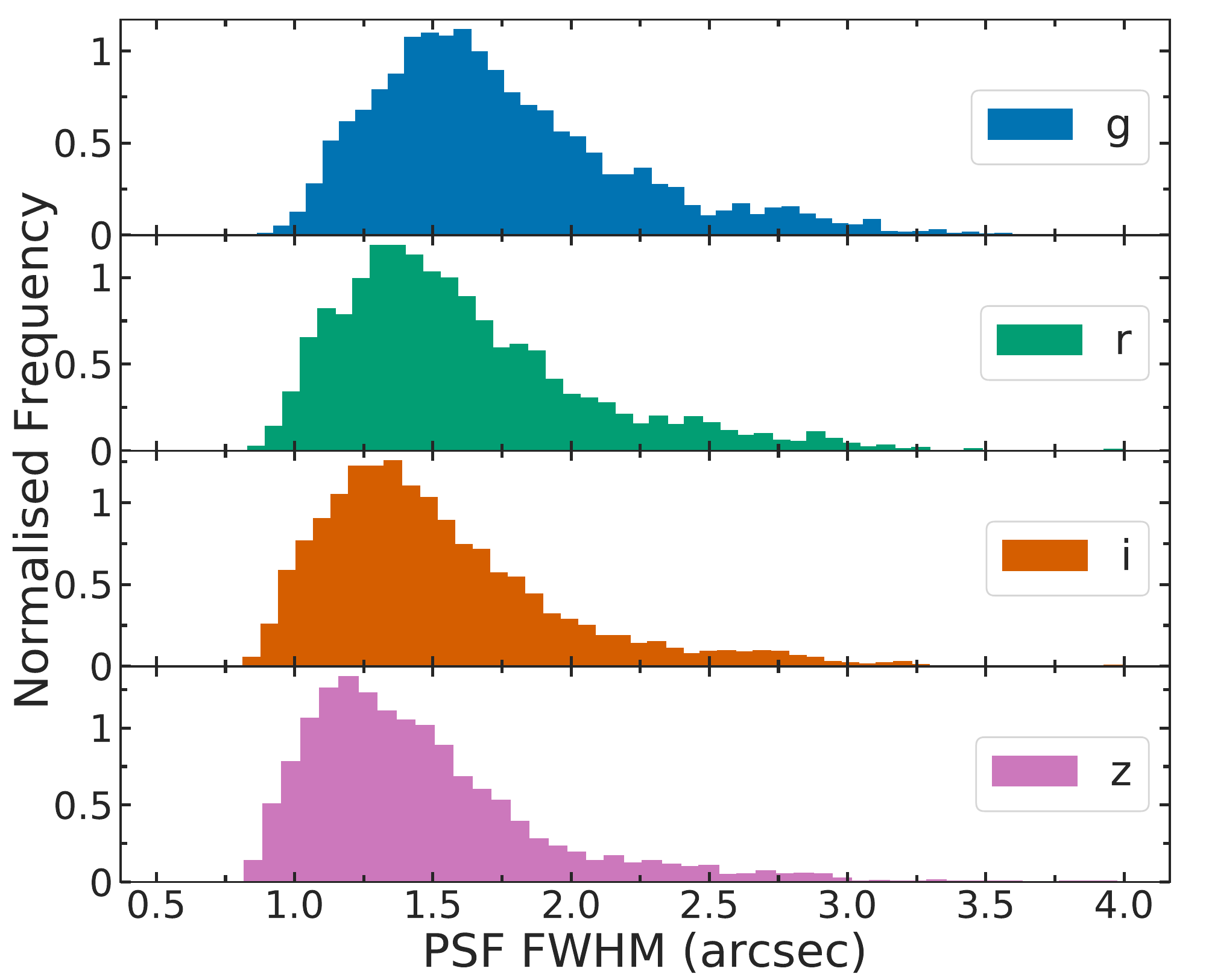}
\caption{Histograms for the distribution of the PSF FWHM across all DES-SN exposures. The distributions share similar shapes with long, high-seeing tails, which are all excluded from the stacks by the seeing cut. The distributions peak at increasingly smaller values as the filter wavelength increases.
\label{fig:psf_hist}}
\end{figure}

\begin{figure}
\includegraphics[width=0.5\textwidth]{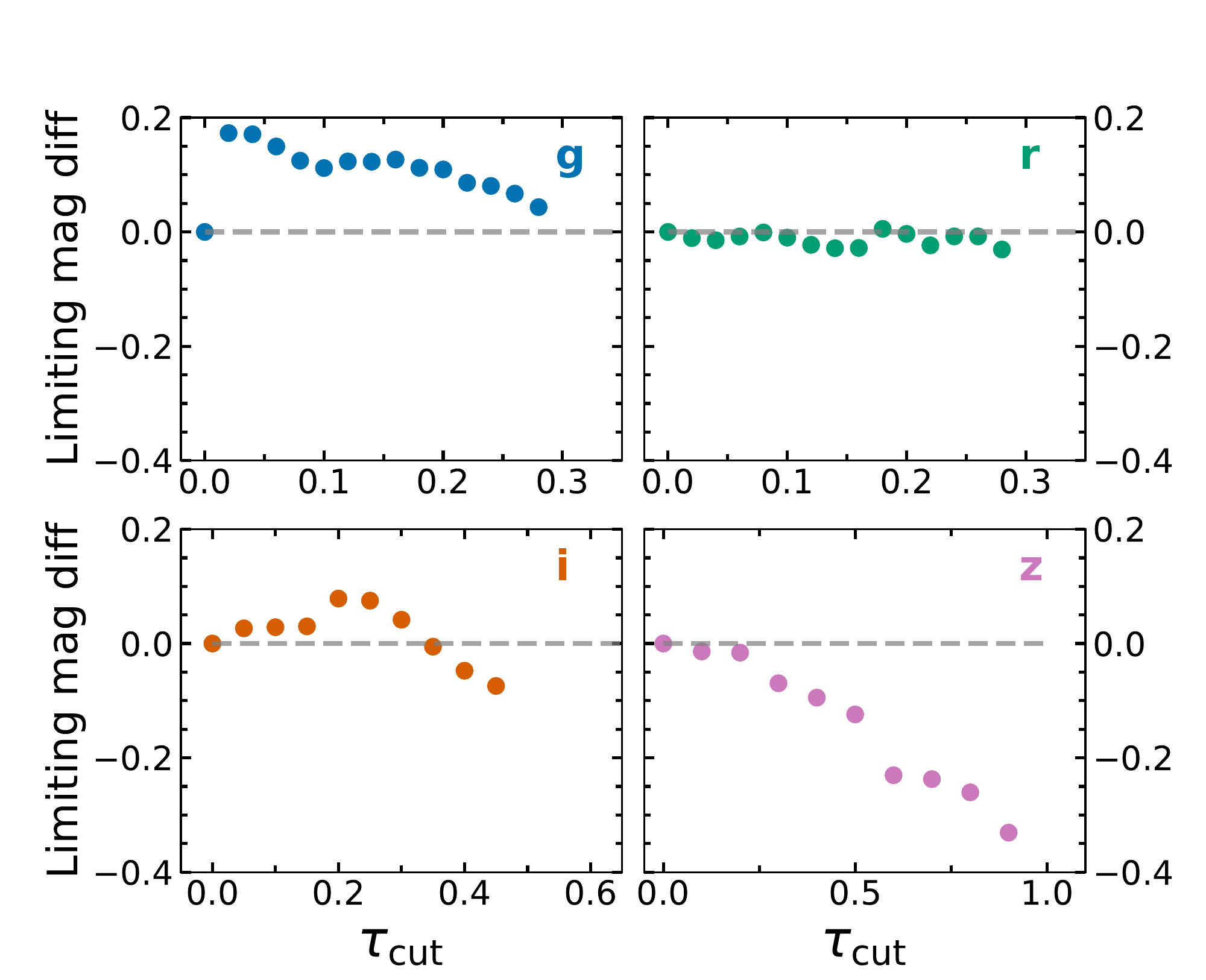}
\caption{The difference in limiting magnitudes (measured from the sky background rms; explained in Section \ref{subsubsec:limmags}) of coadds of CCD 35 in the shallow X2 field, based upon different cuts in $\tau$. The higher the cut in $\tau$, the more single epochs are rejected from inclusion in the stack. The difference is measured compared to the coadd with no cuts ($\tau = 0$). 
\label{fig:optimize_shallow}}
\end{figure}

\begin{figure}
\includegraphics[width=0.5\textwidth]{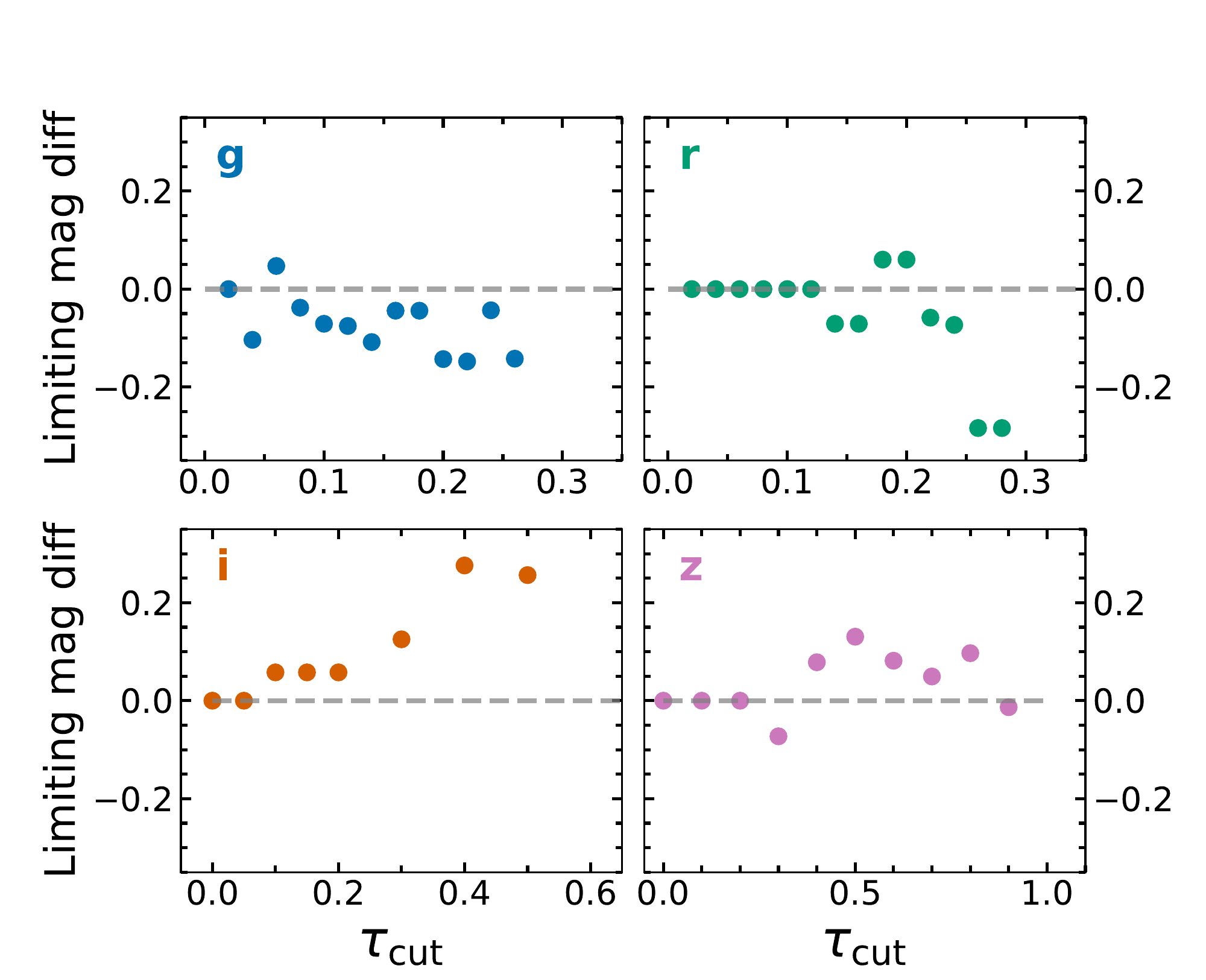}
\caption{As for Fig. \ref{fig:optimize_shallow}, but for CCD 35 in the deep X3 field.}
\label{fig:optimize_deep}
\end{figure}

\section{Deep Photometry}
\label{sec:data}
\subsection{DES-SN survey overview}
DES-SN consisted of a survey of ten separate pointings, grouped into four regions on the sky. Each of these fields was chosen to coincide with a deep extragalactic legacy field: three overlapping with XMM-LSS (the `SN-X' fields; \citealt{Pierre2004}); three with the Chandra Deep Field - South (the `SN-C' fields; \citealt{Xue2011}); two overlapping with the Sloan Digital Sky Survey (SDSS) Stripe 82 (the `SN-S' fields; \citealt{AdelmanMcCarthy2007,Abazajian2009}); and two overlapping with the Elais-S1 field (the `SN-E' fields; \citealt{Oliver2000}). In both the X and C regions, one of the three pointings was subject to longer exposures and is thus denoted a `deep field'. The remaining eight are referred to as `shallow fields'. 
The DES photometric bands, $g$, $r$, $i$, and $z$ correspond closely to their SDSS analogues. Observations were taken in each band roughly every seven days (although exposures failing quality cuts were repeated at the next available opportunity) over five years, comprising six-month observing seasons during consecutive southern summers. Single-visit limiting magnitudes are $m_{\textrm{lim,single epoch}} \sim 23.5$ for shallow fields, and 24.5 for deep fields. Further description of the SN survey and spectroscopic targeting can be found in \citet{Kessler2015,DAndrea2018}. 

\subsection{Coadds\label{subsec:stacks}}

The extensive observations obtained by DES-SN open up the possibility for the creation of deep images by coadding the individual exposures. While the deepest coadds are obtained by combining data from all five seasons, for the purposes of studying SN host galaxies it is important to remove those exposures which contain light from the SN. While it is possible to make an individual coadd for each SN using all exposures minus the exact ones for which that SN was visible, this is computationally expensive and encounters issues such as a precise knowledge of when the SN faded below the detection threshold. As such we create a series of 4-season coadds, for each one excluding all exposures from the other season. We refer to such image stacks as minus-year (MY) coadds: the coadd missing season one is referred to as MY1, and so on. For SNe detected in season $S$, there is a respective coadd MY$S$ for which all exposures from season $S$ are excluded. 
To create the coadds, we use a custom  pipeline\footnote{\url{https://github.com/wisemanp/des\_stacks.git} \label{fn:des_stacks}} that makes extensive use of software from the \texttt{ASTROMATIC}\footnote{\url{www.astromatic.net}} suite. The full set of configuration files used in coaddition and photometry can be found in the publically available \texttt{github} repository in footnote~\ref{fn:des_stacks}.

\subsubsection{Selecting exposures \label{subsec:image_selection}}
The inclusion of particularly poor quality exposures, such as those affected by instrumental noise, high seeing, or clouds, can have a detrimental effect on the quality of a coadd. Poor seeing results in the washing-out of sources in the resulting stacked image, degrading the limiting magnitude for extended sources. While theoretically all epochs with a signal-to-noise ratio greater than unity should improve the depth of the final coadd, empirical tests show that better coadd depth is achieved by introducing selection requirements (cuts), which a single exposure must pass before being included. For this purpose we use the effective exposure time ratio $\tau$ \citep{Neilsen2016}. This is the ratio between the effective exposure time given the conditions, $t_{\mathrm{eff}}$, and the true exposure time $t_{\mathrm{exp}}$, and is given by: 

\begin{equation}
\tau = \eta^2 \left(\frac{\mathrm{FWHM}}{0.9 \mathrm{\arcsec}}\right)^{-2} \left(\frac{b}{b_{\mathrm{dark}}}\right)^{-1}\,,
\label{eq:teff}
\end{equation} where $\eta$ is the atmospheric transmission, $b$ the sky brightness, and FWHM corresponds to the full-width half-maxiumum of the point spread function (PSF) on a particular night. This measure is normalised to the following set of good conditions in the $i$-band: $\eta = 1$, $\mathrm{FWHM} = 0.9 \arcsec$, and $b_{\mathrm{dark}}$ corresponding to the background from a dark sky at zenith. For more details on $\tau$ and $t_{\mathrm{eff}}$ in DES, see \citet{Morganson2018}. 
A value of $\tau = 1$ corresponds to good conditions, while lower values mean that the effective exposure time is shorter than it would have been had the conditions been the same as the fiducial `good' ($\tau=1$) ones. The distribution of $\tau$ over the five years of DES-SN is shown in Fig. \ref{fig:teff_hist}. There is an evident difference in the $\tau$ distributions between filters . The median $\tau$ in $g$ is much smaller than $z$ due to the dependence of atmospheric turbulence on wavelength and the increased degradation caused by the moon at shorter wavelengths. Structure in the histograms, such as steps around $\tau \sim 0.3$, can be explained by the use of data quality thresholds to determine whether observations should be repeated at the next available opportunity \citep{Neilsen2019}. This effect can also be seen in the distribution of seeing measured in the survey in Fig. \ref{fig:psf_hist}, which shows a higher average seeing in the $g$ and $r$ bands. 

To exclude the worst exposures, a $\tau$ cut is made. Exposures for which $\tau$ is below this cut are not included in the coadd. Similarly, we make cuts on the seeing as measured in the initial reduction of the image. In order to find the values for $\tau_{\mathrm{cut}}$ and PSF that optimise the limiting magnitude of the final images, we conduct a series of test coadds. For each test, the exposures passing the corresponding cuts are coadded using the method outlined in Section \ref{subsubsec:coadd_method} and the limiting magnitude is measured (c.f. Section \ref{subsubsec:limmags}) and recorded. Wider ranging $\tau$ cuts and PSF cuts were initially tested, using coarser $\tau$ steps in order to reduce CPU expense. The final optimisation is then run on a smaller range of $\tau$ with finer steps. The optimisation is performed independently in each band, and on a shallow (X2) and a deep (X3) field, although the choice of field does not influence the final adopted cuts.

The optimum PSF cut was found to be 2.4\arcsec ($g$) and 2.2\arcsec ($r$, $i$, $z$) for all fields. Figs. \ref{fig:optimize_shallow} and \ref{fig:optimize_deep} show the results from the $\tau$ optimisation in the shallow and deep field respectively. In some bands, there is a clear evolution in the limiting magnitude based on different $\tau_{\mathrm{cut}}$, although this is not evident in others (e.g. $r$). In the shallow fields, the trend is most obvious in $g$ and $z$, where very lenient cuts, and thus inclusion of all single epochs, result in the deepest coadds. In $r$ and $i$, on the other hand, the depth peaks at $0.2 \leq \tau_{\mathrm{cut}} \leq 0.3$. We note that for $g$, $r$, and $i$, the variation in limiting magnitude between $0 \leq \tau_{\mathrm{cut}} \leq 0.3$ is about $\pm 0.05$ dex, which is smaller than the typical statistical error on objects of such brightness, and also smaller than the RMS variation seen across different CCDs and fields. For reasons related to further optimising the trade-off between depth and computational expense, we chose shallow field $\tau$ cuts of 0.26 in $g$, 0.2 in $r$, and $i$ and 0.3 in $z$. For the same reasons, in the deep fields we choose $\tau$ cuts of 0.06, 0.2, 0.4 and 0.5 for $g$, $r$, $i$, and $z$ respectively. We note that the $i$-band limiting magnitude increases with $\tau_{\mathrm{cut}}$ in the deep fields, whereas the other bands are relatively flat. Similarly, the variation in the deep field $\tau_{\mathrm{cut}}$ values is larger than for shallow fields. We suggest it is likely caused by the differing distributions of $\tau$ in each filter although it is not immediately clear why this should be the case.
We stress that the results presented in this and future analyses are robust to small shifts in the $\tau$ cut, as the inclusion/exclusion of single images has a negligible affect on whether a host is detected or not. A summary of the cuts for each field and band is given in Table \ref{tab:coadds}.

While here we optimise the stacks for their ultimate depth, using limiting magnitude as a diagnostic, this simple method makes it possible to quickly optimise the cutting procedure for any desired output variable. For example, the analysis of \citet{Kelsey2020} uses a version of these coadds that has been optimised for the best seeing in order to resolve sub-galactic scale regions of SN Ia hosts, to improve measurements of local properties around the SN locations. 

\subsubsection{Coaddition \label{subsubsec:coadd_method}}

Individual exposures are detrended through the Dark Energy Survey Image Processing Pipeline \citep{Morganson2018}. To stack the individual exposures, we use \texttt{SWarp} \citep{Bertin2002}. Each of the 59 DECam science CCDs functioning for the entire survey duration are treated independently. For each chip, in each field, band, and MY combination all exposures passing the relevant PSF and $\tau$ criteria are resampled using the default Lanczos-3 6 $\times$ 6-tap filter, and then coadded. The resampling may affect photometric uncertanties by introducing correlations between resampled pixels. However, this effect is negligible compared to the dominant zeropoint uncertainty (Section \ref{subsec:phot}). Due to the large number of input exposures, which themselves are already deep, the commonly-used mean and weighted mean stacking methods lead to the contamination of the final coadd by a high density of artefacts such as satellite trails and cosmic rays. Median and clipped median stacks, which are efficient at removing artefacts, lead to systemetic offsets in the photometry of bright objects \citep[e.g.][]{Gruen2014}, due to the inhomogeneity of the PSFs of the single epochs. We therefore utilise the clipped mean stacking method (\citealt{Gruen2014}, and code therein\footnote{\url{https://web.stanford.edu/~dgruen/download.html}}), whereby outlier pixels are detected by performing a clipping procedure. The detected outlier regions in individual exposures are masked, before the implementation of a weighted average stack, using inverse variance weight maps. This method has previously been implemented in several analyses \citep[e.g.][]{Melchior2015,Gruen2014a}.



\begin{figure}
\includegraphics[width=0.5\textwidth]{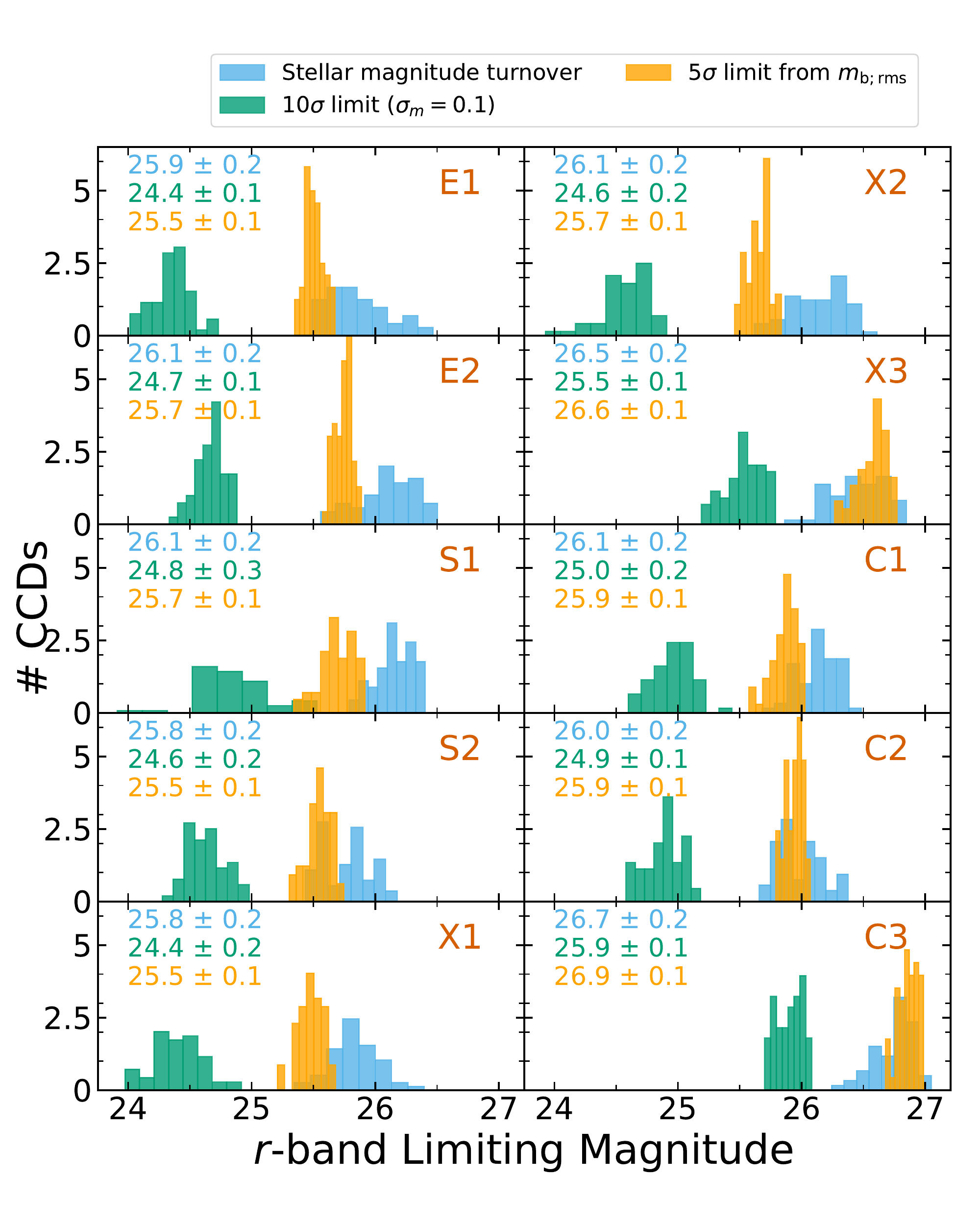}
\caption{Histograms of the limiting $r$-band magnitudes for all CCDs in each of the ten SN fields for the MY1 coadd. The extra magnitude of depth in the deeper fields, X3 and C3, is clearly evident. The means and standard deviations of the limiting magnitude distributions are displayed in the upper left corner for each field. While we report $5\sigma$ limits in this catalogue, we also show $10\sigma$ limits from magnitude uncertainties (green) for consistency with previous catalogues. Summaries of the coadd properties are given in Table \ref{tab:coadds}.
\label{fig:limmags}}
\end{figure}

\begin{figure}
\includegraphics[width=0.5\textwidth]{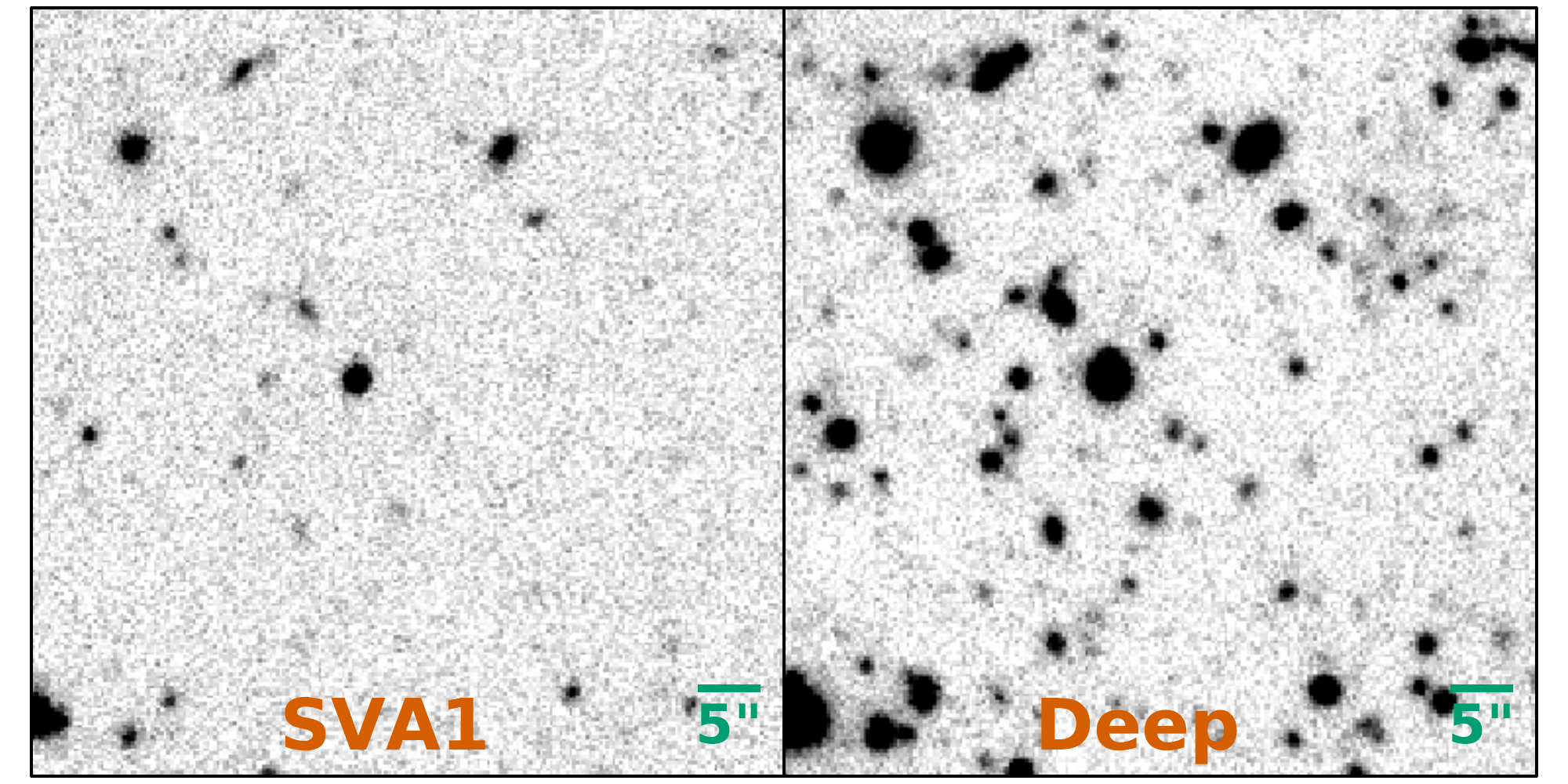}
\caption{A comparison between the $r$-band coadds from SVA1 (left), and this work (right), for a small region of the C3 field. The addition of detail on brighter objects as well as the detection of fainter objects is clearly visible.
\label{fig:stamp_eg}}
\end{figure}

\subsection{Photometry\label{subsec:phot}}

\subsubsection{Calibration \label{subsubsec:calib}}
To perform a photometric calibration on the coadds, we calculate photometric zeropoints by matching stars to existing catalogues. The zeropoints are then used to calibrate the common aperture photometry described in Section \ref{subsubsec:cap}. Sources for use in the calibration are detected using \texttt{Source Extractor} \citep{Bertin1996}. In order to calculate zeropoints for the coadd images, detected sources are matched to a catalog made using the first three years of the DES wide-area survey, known as Y3A1, via the DES Image Processing Pipeline \citep{Morganson2018}. We make use of the \texttt{MODEST} classifier \citep[e.g.][]{Chang2015} to select robustly classified stars from the catalog, imposing the additional criterion that stars must be brighter than 22nd magnitude, where the scatter is lowest and the \texttt{MODEST} classifications are most robust. We exclude stars brighter than 18th magnitude, as we find that the coaddition technique leads in some cases to the clipping of the centres of the images of particularly bright stars. We calculate the magnitude zeropoint and its uncertainty for each deep image by using the median of the zeropoints from each individual bright star match, and the corresponding median absolute deviation (MAD) divided by the square root of the number of stars: $\mathrm{ZP_{err}} = \mathrm{MAD}/\sqrt{n}$. This uncertainty dominates the total photometric uncertainty, particularly for brighter objects whose statistical uncertainty is lower.  

\begin{table*}
\caption{Summary of the median differences between magnitudes (errors) of matched objects in the deep coadds from this work compared to those in SVA1 and Y3A2\_DEEP. \label{tab:coadd_qa}}
\begin{tabular}{lrrrrrrrr}
\toprule
Comparison &     $\Delta m_g$ &  $\Delta m_g$ ($m_g>24$) &     $\Delta m_r$ &  $\Delta m_r$ ($m_r>24$) &     $\Delta m_i$ &  $\Delta m_i$ ($m_i>24$) &     $\Delta m_z$ &  $\Delta m_z$ ($m_z>24$) \\
\midrule
SVA1 mag           &  0.06 &     0.00 &  0.03 &     0.03 & -0.06 &     0.09 &  0.02 &     0.10 \\
SVA1 mag err       & -0.09 &    -0.12 & -0.06 &    -0.09 & -0.05 &    -0.09 & -0.06 &    -0.13 \\
Y3A2\_Deep mag     &  0.05 &     0.02 &  0.05 &     0.12 & -0.05 &     0.12 & -0.04 &     0.14 \\
Y3A2\_Deep mag err & -0.02 &    -0.03 & -0.03 &    -0.04 & -0.03 &    -0.05 & -0.06 &    -0.12 \\
\bottomrule
\end{tabular}
\end{table*}

\begin{figure}
\includegraphics[width=0.5\textwidth]{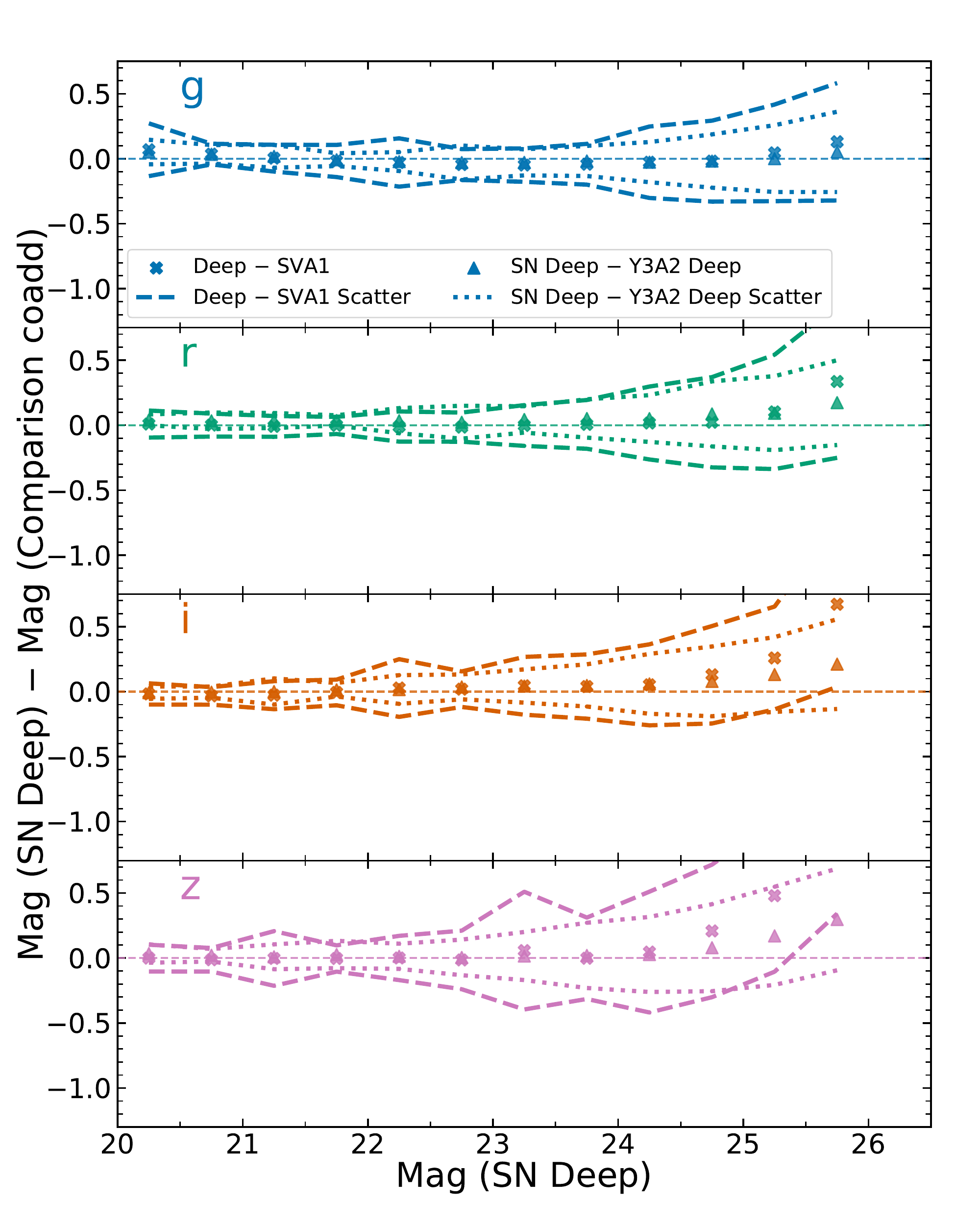}
\caption{The difference in magnitude between the SN Deep coadd (this work) and two comparison catalogues: the SVA1 coadd, and the Y3A2 Deep coadd. Shown here are the differences for objects that lie in the X2 field, binned by their brightness. The dashed and dotted lines trace the scatter in the magnitude differences for each comparison respectively. In general, the data is centred around a magnitude difference of 0, indicating that the photometry is consistent. 
\label{fig:magdiff_combined}}
\end{figure}

\begin{figure}
\includegraphics[width=0.5\textwidth]{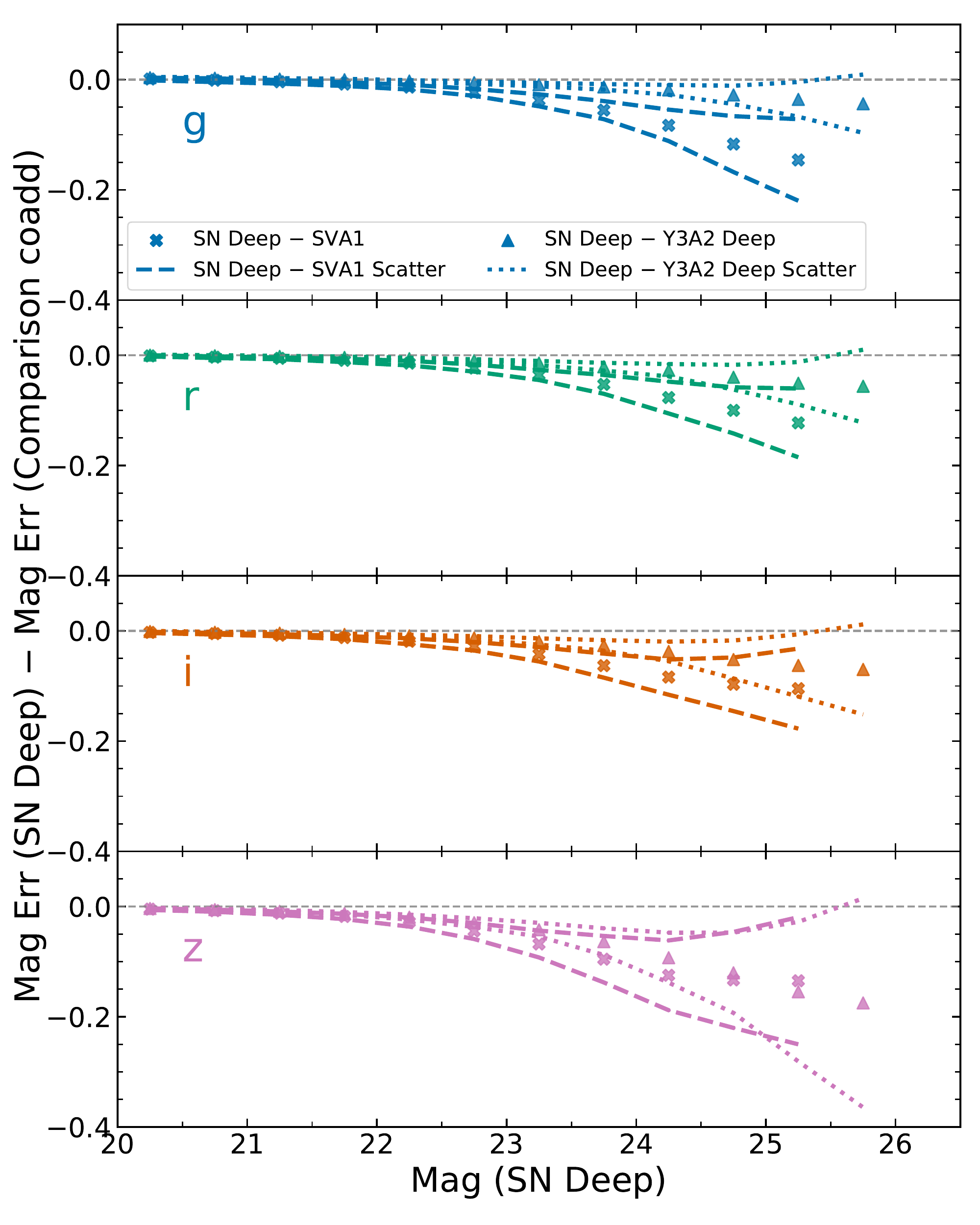}
\caption{As per Fig. \ref{fig:magdiff_combined}, but comparing the differences in magnitude errors (Mag Err). The increasingly negative tail at fainter magnitudes implies more precisely measured values in SN Deep.}
\label{fig:magerrdiff_combined}
\end{figure}

\begin{figure}
\includegraphics[width=0.5\textwidth]{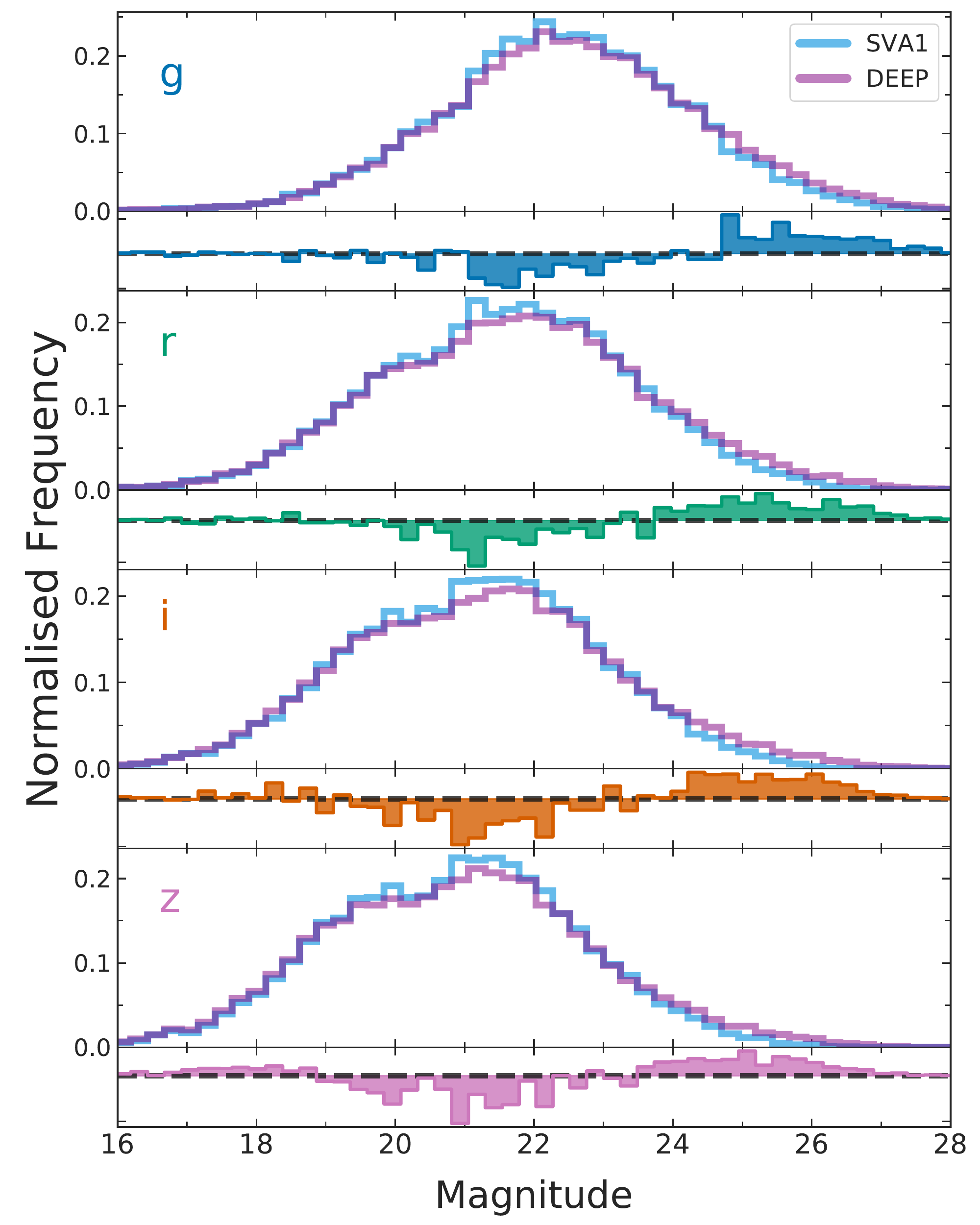}
\caption{The normalised histograms of the magnitudes of SN host galaxies in the SVA1 (cyan) and the deep (purple) catalogues (upper panels), and deep $-$ SVA1 (lower panels). At magnitudes where the difference is negative, a larger fraction of the SVA1 hosts lie at that magnitude than the corresponding fraction in the deep catalogue. Conversely, positive differences mean there are the fraction of deep hosts at that magnitude is greater than the fraction of SVA1 hosts. In all four bands, the distributions are skewed more strongly to fainter magnitudes. 
\label{fig:compare_mag_dists}}
\end{figure}

\subsubsection{Limiting Magnitudes \label{subsubsec:limmags}}
There are multiple methods to calculate the limiting magnitude of an image; that is, the magnitude fainter than which limits are reported rather than detections. Firstly, the limiting magnitude can be approximated from the distribution of the magnitudes of detected sources. The magnitude at which the distribution peaks is taken as the limiting magnitude. This is because the true magnitude distribution of sources rises to much fainter values, so the turn off is indicating that some objects are not being detected. This method is quite strongly dependent upon the parameters used during the source extraction process such as the detection threshold chosen, since using a lower detection threshold will push the peak of the distribution to fainter magnitudes, but there is a greater chance that these detections will be false. 

Secondly, a limiting magnitude can be calculated using the measurement uncertainty of the magnitudes of detected objects. The limit is simply the magnitude at which the mean magnitude error of objects, $\sigma_m$, is equal to a threshold determined by the precision required. $10\sigma$ is the typically quoted value in DES \citep[e.g.][]{Jarvis2015}, and thus our threshold becomes $\sigma_m = 0.1$.

Thirdly, one can calculate the limiting magnitude using the rms of the background, $m_{\mathrm{b;rms}}$, using:

\begin{equation}
\label{eq:limmag_sky}
m_{\mathrm{lim}} = m_{\mathrm{zp}} - 2.5\log\left(n\times m_{\mathrm{b;rms}}\sqrt{\pi \mathrm{FWHM}^2}\right)\,,
\end{equation}
where $n$ corresponds to the sigma-level required, and $ m_{\mathrm{zp}}$ is the zeropoint magnitude. For this catalogue we report $5\sigma$ limits. The FWHM is the mean measured for the point sources in the field, in pixels. While the first two methods use detected objects (and thus includes both stars and galaxies), this method use the area of the aperture used for photometry. Here we use the median FWHM of all objects in order to calculate an average object detection limit.

The distributions of limiting magnitudes measured in the above ways is shown in Fig \ref{fig:limmags}. The $5\sigma$ sky magnitude limit and stellar magnitude turnover are broadly consistent with each other for all fields. The $10\sigma$ measurement based off magnitude uncertainties is brighter, as expected, than the $5\sigma$ limit from the background. The deep fields, X3 and C3, are a magnitude deeper in all three diagnostics. We note that C3 is deeper than X3, due to $\sim 30$ more epochs, corresponding to $\sim 3$ hours, passing cuts (Table \ref{tab:coadds}).
For the assessment of depth used in Section \ref{subsec:image_selection}, we use the sky magnitude, as it is independent of the choice of source-detection parameters. During the initial set of tests, we trialled using different measurements of depth but the effect on the chosen cuts was negligible.

\subsubsection{Common Aperture Photometry \label{subsubsec:cap}}

The most accurate photometry requires a model of the PSF (for point sources) as well as a morphological model with which to convolve it (for galaxies). In \texttt{Source Extractor} these techniques correspond to the \texttt{MAG\_PSF} and \texttt{MAG\_MODEL} magnitudes, respectively. However, in the deep coadds the PSF of the final image is a combination of the PSFs of the  of $\sim 100$ individual exposures. Such a composite PSF is non-trivial to model and as such renders those magnitude measurements unreliable, although efforts have been made to homogenise the PSF at the coadding stage \citep{Mohr2012}. Instead, as we are chiefly interested in fitting galaxy SEDs, we employ common aperture photometry (CAP) in order to ensure we are detecting light from the same physical area in each band, and therefore maintaining consistent galaxy colours. For CAP, \texttt{Source Extractor} is run in dual-image mode, whereby the measurement apertures are defined on a detection image and used for the subsequent measurement in all four bands. For the detection image, we use a simple average combination of $r+i+z$. We also trialed using $g+r+i$, $g+r+i+z$, as well as just $i$ and $z$ as detection images, and found that $r+i+z$ is most reliable at detecting faint objects. The magnitudes recovered when using different detection images are consistent within the measurement uncertainties. \\
One of the biggest issues encountered in source detection in deep images is the vast dynamic range of source brightnesses that we wish to measure. Galaxies in the deep fields span the magnitude range $14 \leq M_r \leq 28$. Naturally, we wish to report magnitudes that are accurate across this range, but particularly at the fainter end, where the existing catalogs we seek to improve on do not extend to. A large part of the problem in detecting and measuring faint objects is the task of deblending, where they may lie close to, or overlap with (in either a physical or projected sense) a much larger, brighter source. The detection of such objects can be achieved by tuning the detection parameters in \texttt{Source Extractor}. The parameters refer to flux threshold (compared to the background), and number of pixels above this threshold, required for a detection. We set these low at $1.25\sigma$ and 3 pixels respectively, such that objects can be PSF-size and of low significance to count as detections. While raising the number of false detections, the number of these is small upon visual inspection. The low thresholds allow small, faint objects to be detected but they are often located close to larger brigher objects and \texttt{Source Extractor} may deem them to be part of the same object. Using \texttt{Source Extractor} default values for deblending parameters in an initial run, we compared the recovered host galaxies of DES-SNe with the corresponding hosts in SVA1. Four SN host galaxies with detections in SVA1 had not been picked up in the deep coadds, of which all were in small, faint, PSF-like sources near large, bright galaxies. We adjusted the deblending parameters\footnote{The deblending procedure is explained in detail in the unofficial \texttt{SExtractor} manual \url{http://astroa.physics.metu.edu.tr/MANUALS/sextractor/Guide2source\_extractor.pdf}} until these hosts were detected correctly.

We use the Kron magnitude (\texttt{MAG\_AUTO}) output from \texttt{Source Extractor}, as well as circular aperture measurements with the aperture diameter set to 2\arcsec. This diameter corresponds to the width of the AAOmega fibres to allow for direct comparisons with, and calibrations of, galaxy and transient spectra taken as part of the DES-SN spectroscopic follow-up programme, OzDES 
(\citealt{Yuan2015,Childress2017}, Lidman et al. \textit{in prep}).

We use the magnitude zeropoints previously calculated from the good quality stars to calibrate the common aperture photometry, resulting in the catalogue we name SN Deep.

%

\subsection{Performance \label{subsec:qa}}
A section of the SN Deep coadd is shown in Fig. \ref{fig:stamp_eg} and is compared to the SVA1 coadd. The increased depth is evident due to the large number of extra sources detected, as well as the extent of existing objects.

The SVA1 $r$-band $10\sigma$ limiting magnitude is reported as approximately 23.8\footnote{\url{https://des.ncsa.illinois.edu/releases/sva1}}. In Fig. \ref{fig:limmags}, the green histograms reveal the SN Deep limiting magnitudes to be on average between 0.6 and 1.2 magnitudes deeper in the shallow fields, and 1.7 to 2.1 magnitudes deeper in the deep fields. In the $z$-band, the difference is closer to 2 magnitudes in the shallow fields and 3 magnitudes in the deep fields, demonstrating the relative enhancement at redder wavelengths.

To assess the quality of the deep photometry, we compare our photometry to previous DES catalogues of the SN fields. In this section, results will be listed corresponding to the $g$, $r$, $i$, and $z$ bands respectively. The median differences in magnitude and uncertainty for all objects, as well as those fainter than 24th mag, are presented in Table \ref{tab:coadd_qa}. In Fig. \ref{fig:magdiff_combined}, we plot the residual between the magnitudes from a subset of objects in a single CCD of our deep catalogues and the same matched objects in SVA1 as well as a set of deep coadds using a different processing pipeline, known as Y3A2\_DEEP (DES Collaboration, \textit{in prep.}). The median offsets between the catalogues are presented in Table \ref{tab:coadd_qa}, with values quoted for all sources, as well as for faint sources with mag $> 24$. The offsets between SN Deep and the two comparison catalogues are small, with the absolute mean differences to Y3A2\_Deep at 0.05 mag or smaller. There is a general trend towards a positive offset at the fainter (i.e. $>25$) magnitudes, with SN Deep reporting around 0.1 mags fainter than Y3A2\_Deep (the comparison to SVA1 at $m>25$ is uninformative as it is beyond the typical depth of that catalogue). The scatter in the differences, traced by the dashed lines, increases with magnitude. The scatter is smaller for the comparison to Y3A2\_Deep than for SVA1, reinforcing the assumption that magnitudes reported in the deeper catalogues are closer to the `truth' values. 

Fig. \ref{fig:magerrdiff_combined} shows the difference in the magnitude uncertainties (\texttt{MAGERR\_AUTO}) between the same matched galaxies. In addition to the statistical uncertainty, we also include the zeropoint uncertainty in our final magnitude uncertainties. The uncertainty reported from SVA1 is systematically larger than the combined SN Deep uncertainty, and the difference increases at fainter magtniudes. This trend also exists in the comparison with Y3A2\_Deep, although the strength is band-dependent, with more improvement noticeable with increasing wavelength. 

\subsection{Host matching \label{subsec:host matching}}
To remain consistent with the DES-SN3YR method \citep{DAndrea2018}, we match transients to host galaxies in SN Deep using the directional light radius (DLR) method \citep{Sullivan2006,Gupta2016}. The matching algorithm chooses the host with the smallest $d_{\mathrm{DLR}}$, which is the ratio between angular separation between the transient and the centre of a galaxy, and the size of that galaxy in the direction of the transient. As with \citet{DAndrea2018} we use a threshold of $d_{\mathrm{DLR}}\leq 4$; any object with no galaxies within this threshold is determined to be hostless. 

\subsubsection{Changes from SVA1\label{subsubsec:sva1-deep changes}}
In total, of the 31473 transient candidates in DES-SN, 24695 (78.5\%) have an assigned host galaxy in the SVA1 catalogue. Using the deep coadds, that number is increased to 27548 (87.5\%). Of the transients with SVA1 hosts, 23943 (97\%) have the same host in SN Deep, with 280 objects (that is, 1.1\% of those that already had a host) changing to a different host. A further 3325 (10.6\% of all transients) have a host in SN Deep that was not the host in SVA1; for 547 (1.7\% of all transients) of these, the SN Deep host was listed in the SVA1 catalogue but had $d_{\mathrm{DLR}}>4$ and thus was not considered the host. 26 objects move the other way - that is, their SVA1 host is the best match in SN Deep, but is now $d_{\mathrm{DLR}}>4$. Finally, there are 446 (1.4\%) transients that had a host in SVA1 but that galaxy is not detected in SN Deep. These objects are located in the gaps between CCDs in the deep coadds, a problem which was avoided in SVA1 by tiling observations. For these objects, we use the SVA1 data in the SN Deep catalogue. For the spectroscopically confirmed sample of SNe Ia used in the DES3YR analysis, the use of SN Deep means 13 SNe Ia (6.3\%) are assigned a host when they did not have one in SVA1, while 4 objects (2\%) are assigned a different host. 1 object lies on a chip gap and thus is not covered by SN Deep. A comparison of the behaviour of the SN Ia host galaxy mass step with the use of this photometry versus that from SVA1 can be found in Smith et al., \textit{in prep} (Hereafter S19).

As a further measure of the increased depth of the deep coadds, we calculated the apparent magnitude distributions of the assigned hosts of all DES transients using the DLR method in both SVA1 and SN Deep, and show the results in Fig. \ref{fig:compare_mag_dists}. The difference between the two distributions are shown beneath for each band. The shape (negative at bright magnitudes, positive at faint magnitudes) is caused by transients whose assigned host has changed from a brighter galaxy in SVA1 to a fainter galaxy in SN Deep. The transition appears roughly consistent with $g = 24.5$, and at increasingly brighter magnitudes through the longer wavelengths, which corresponding to the evolution of the magnitude limit of SVA1. 
A thorough exploration of how host mismatching may affect cosmological studies will be presented in a future paper following the technique of \citet{Popovic2019}.

\section{Host SED Modelling }
\label{sec:seds}
In this section, we estimate stellar masses for the host galaxies of various sub-samples of DES-SN transients. The goal is to be able to compare the host stellar mass distribution of various classes of transients to those observed in other surveys. The DES and comparison samples are introduced in the respective sections below. The true power of the SN Deep catalogue will eventually lie in the analysis of host galaxies of large, photometrically selected transient samples from DES, which are to be presented in the near future and are beyond the scope of this paper. As such, here we use predominantly spectroscopically selected samples and choose a variety of literature samples with which to compare our data. We do so in a proof of concept fashion, to showcase both the precision in mass measurements made possible by the coadds, as well as to introduce a probabilistic method to compare samples.

\subsection{Parameter Estimation}

In order to estimate stellar masses of the galaxies comprising the DES-SN host samples, we fit the $griz$ SED of each host galaxy with templates formed of a combination of simple stellar population models. The fitting method, which makes use of the \texttt{P\'EGASE.2} spectral synthesis templates \citep{Fioc1997,LeBorgne2002} and a \citet{Kroupa2001} initial mass function (IMF), is described in \citetalias{Smith2020}). As per \citet{Palmese2019}, we include a 0.1~dex systematic uncertainty on derived masses due to an apparent degeneracy between stellar mass and dust extinction.

\subsection{Host galaxy samples \label{subsec:sample}}

Below, we describe the host galaxy samples for which we calculate stellar masses. The SN selection and classifications used here are described primarily in \citet{DAndrea2018}.

The host galaxy samples are derived by matching each transient to galaxies detected in the deep coadds via the DLR method outlined in Section \ref{subsec:host matching}. To be included in the following analysis, the host galaxy must have an associated redshift, provided either by the dedicated DES spectral follow-up programme OzDES at the Anglo-Australian Telescope (AAT) with redshift flag 3 or 4 (see \citealt{Childress2017} for details on OzDES redshift flags), other legacy redshift catalogues, or derived from the classification spectra of the SN itself. The requirement of having a spectroscopic redshift introduces various selection biases to the samples that are not well characterised. In detail, the selection function will depend not only on the brightness of the hosts, but also on the strength of emission and absorption features in the host spectra \citep{Yuan2015}. An exploration of the implications of the host galaxy selection function will be presented in M\"oller et al. \textit{in prep}.

\begin{figure}
\includegraphics[width=0.5\textwidth]{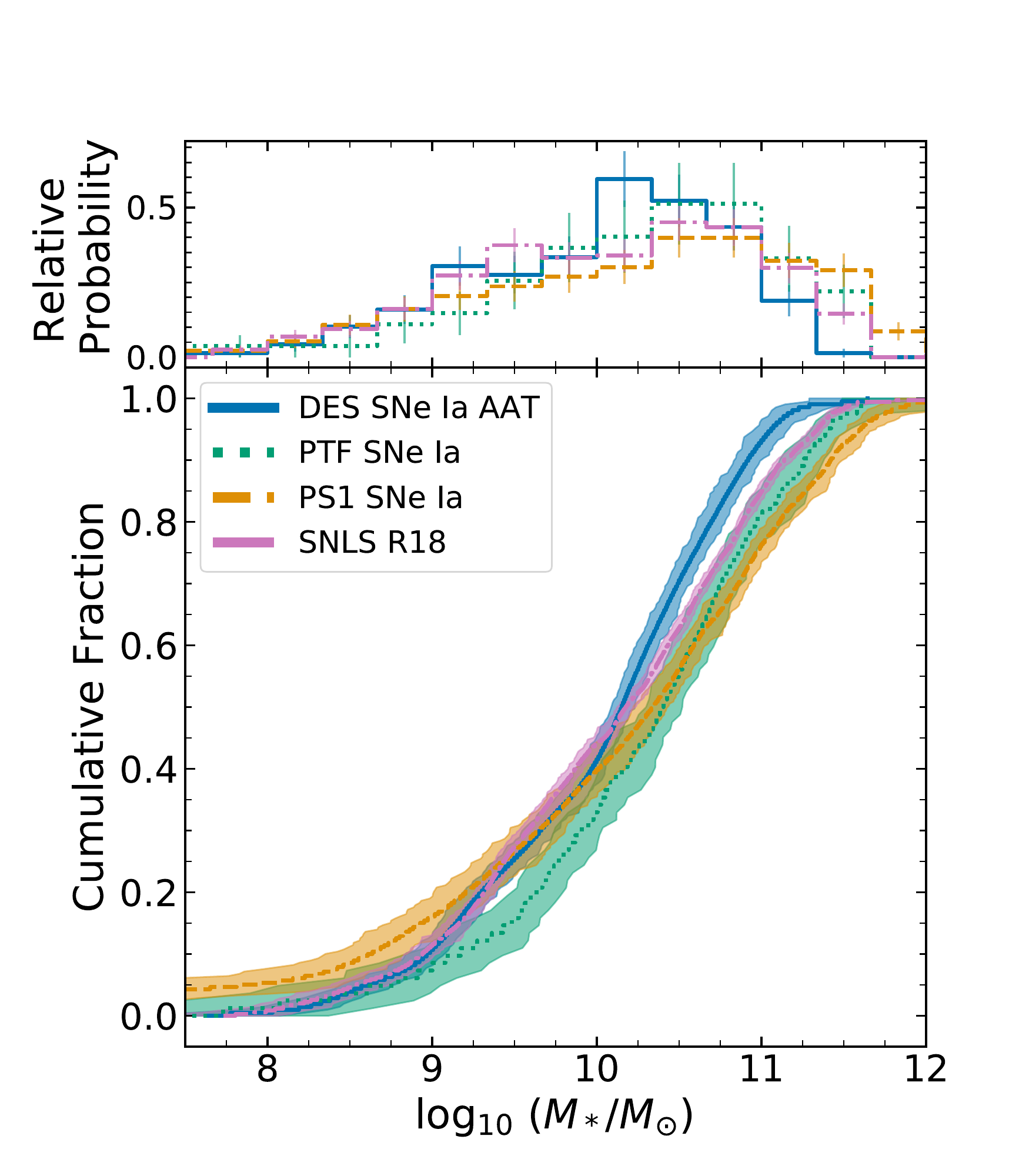}
\caption{The cumulative distribution of the stellar masses of SNe Ia hosts in DES (this work \& \citetalias{Smith2020}), PS1 \citep{Scolnic2018}, PTF \citep{Pan2014}, and SNLS \citep{Roman2018}. The CDF has been estimated using a Monte Carlo oriented technique similar to survival analysis. The shaded regions represent the $1\sigma$ uncertainties on the distribution (Section \ref{subsec:res_Ia}). DES separates from the other samples in the $10^{10}$ - $10^{11}~M_{\sun}$ mass range.
\label{fig:Ia_cum}}
\end{figure}

\begin{figure}
\includegraphics[width=0.5\textwidth]{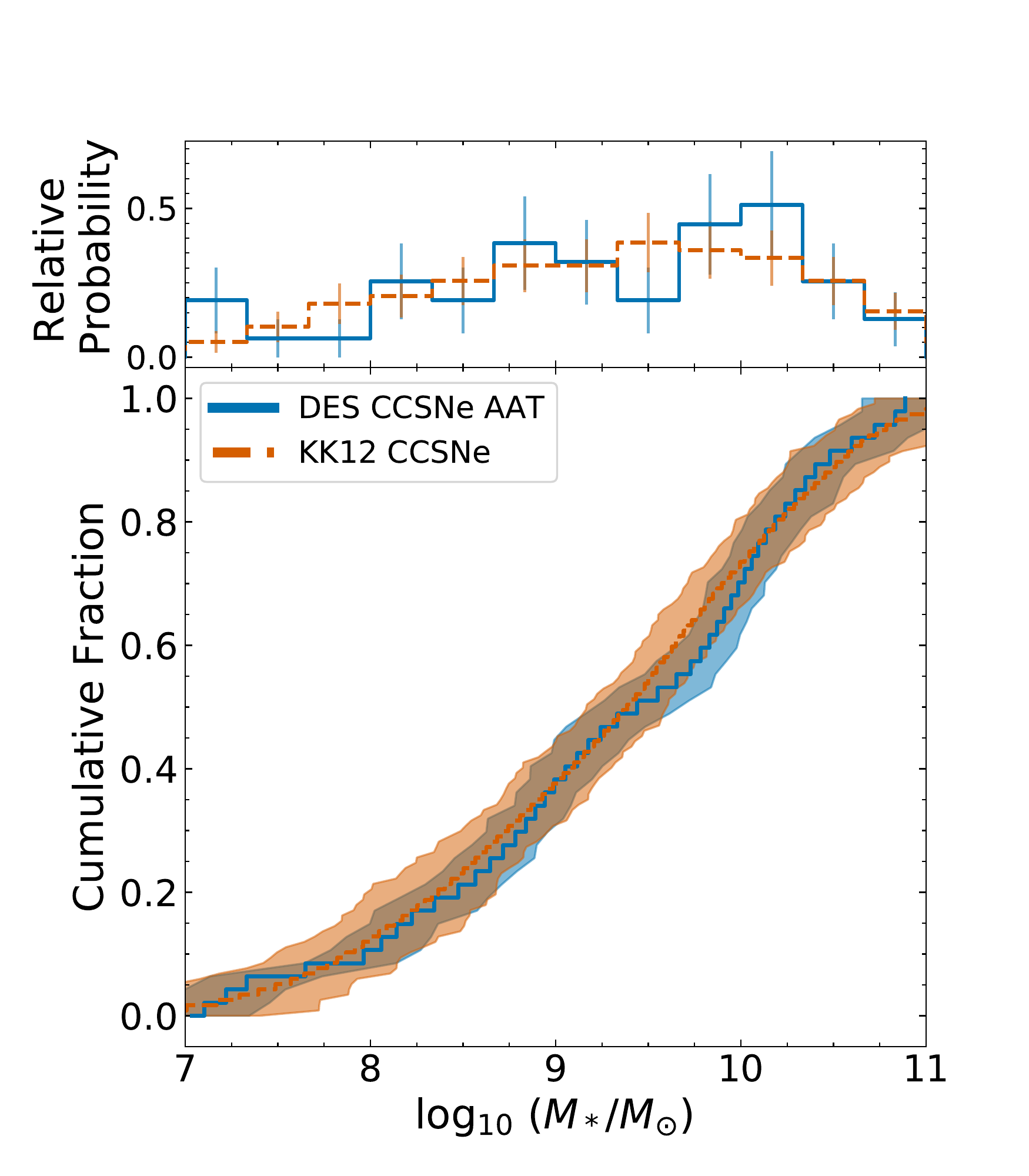}
\caption{CCSNe host stellar mass distributions from DES (AAT) (this work) and \citetalias{Kelly2012}. The distributions appear consistent with one another.
\label{fig:cc_cum}}
\end{figure}

\begin{figure}
\includegraphics[width=0.5\textwidth]{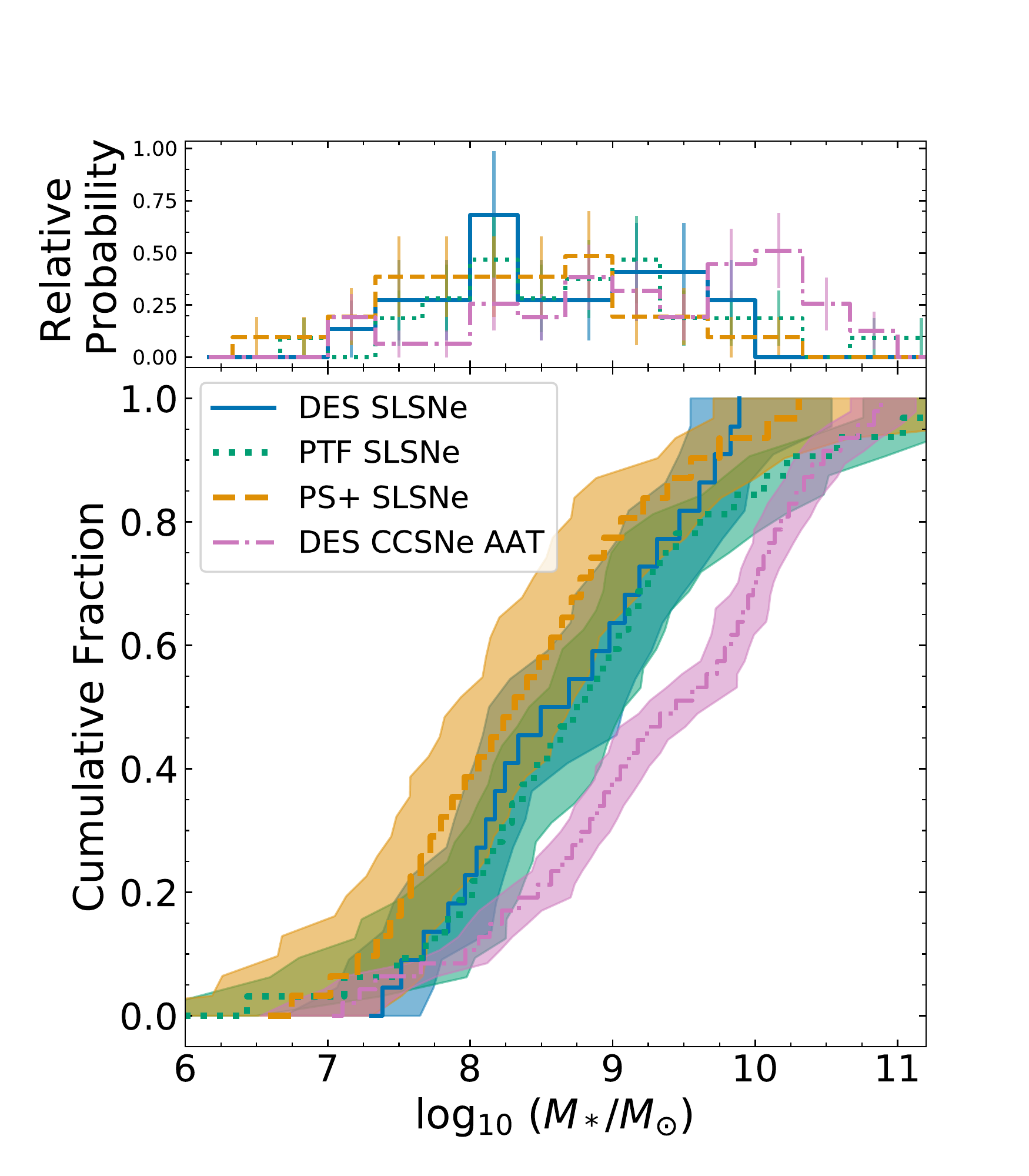}
\caption{SLSN host stellar mass distributions from DES (this work \& \citealt{Angus2019}), PTF \citep{Perley2016a}, PS+ \citep{Lunnan2014}, as well as DES CCSNe (this work). SLSNe hosts are systematically less massive than CCSNe. The DES sample is consistent with both literature samples.}
\label{fig:slsn_cum}
\end{figure}

\begin{figure}
\includegraphics[width=0.5\textwidth]{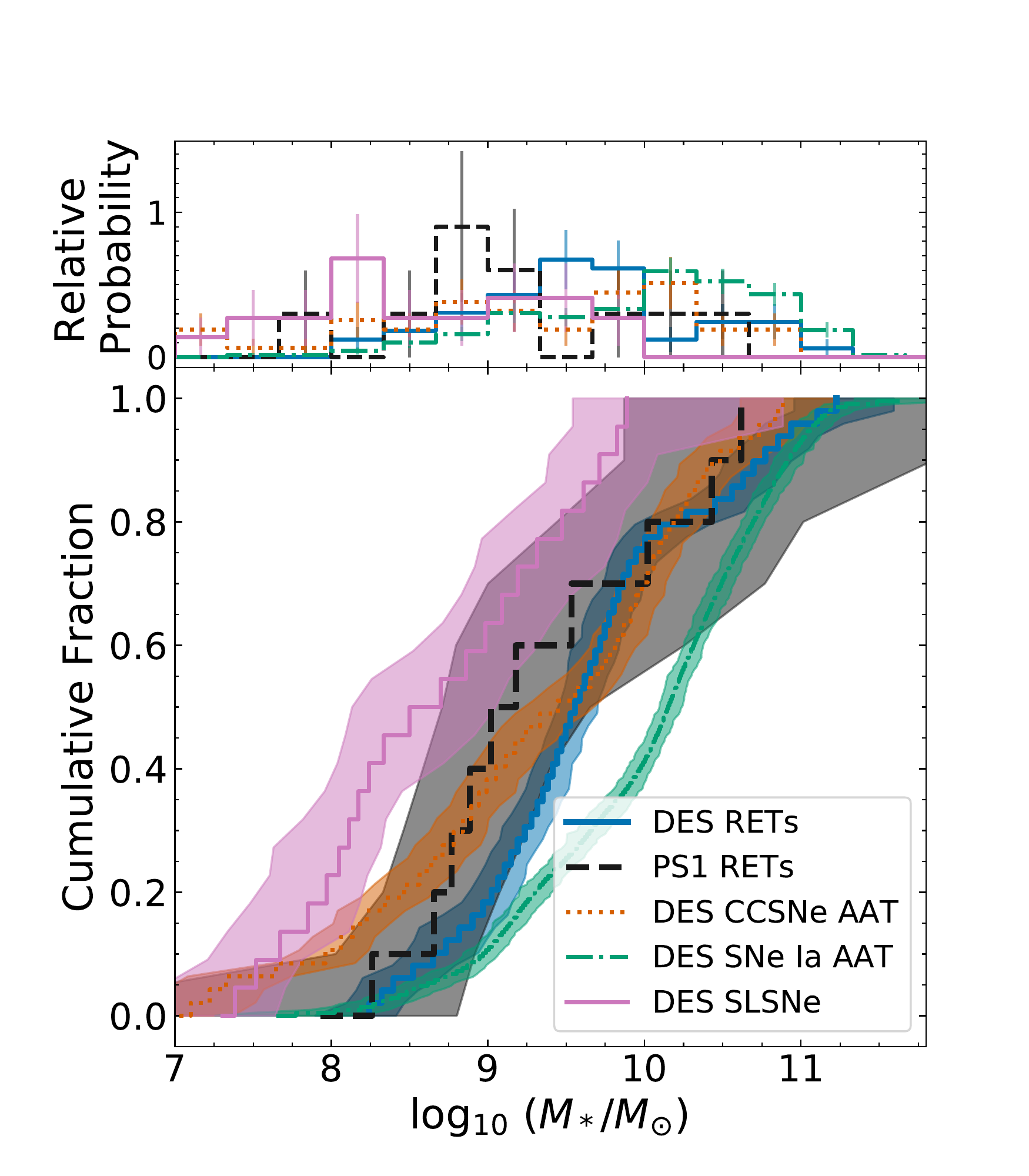}

\caption{RET host stellar mass distributions from DES (this work) and PS1 \citep{Drout2014}, as well as CCSNe, SNe Ia, and SLSNe from DES (this work).}
\label{fig:ret_cum}
\end{figure}

Here we describe the DES subsamples along with the selection of literature samples we use for comparison.

\subsubsection{SNe Ia \label{subsubsec:sample_Ias}}
We model the host galaxy SEDs for spectroscopically confirmed SNe Ia from the full five years of DES-SN. We include SNe Ia with DES classifications \texttt{SNIa} and \texttt{SNIa?} as defined in \citet{DAndrea2018}, which are the classifications used in the DES3YR cosmological analysis. Classifications were obtained with a number of different telescopes and instruments under different programmes, including one to specifically target SNe in faint host galaxies. In order to minimise host-galaxy selection bias, we include only SNe classified by the magnitude-limited OzDES live-transient follow-up programme \citep{DAndrea2018}, and thus refer to the sample as DES SNe Ia (AAT). We further impose the restriction that the host galaxy must have a measured redshift. This redshift requirement means we exclude `hostless' objects, thus likely biasing our sample to higher masses. This sample comprises 207 galaxies with mean redshift $\hat{z} = 0.30$. 

We compare the DES (AAT) sample to the host galaxies of 82 SNe Ia in the local Universe from the Palomar Transient Factory (PTF; \citealt{Pan2014}) with $\hat{z} = 0.05$, as well as 279 at cosmological distances in the PanSTARRS 1 survey (PS1; \citealt{Scolnic2018}) with $\hat{z} = 0.29 $ and 353 from the Supernova Legacy Survey (SNLS) using the sample of \citet{Roman2018} with $\hat{z} = 0.63$. The stellar masses in the above samples have all been measured using \texttt{P\'EGASE.2} templates.

\subsubsection{CCSNe \label{subsubsec:sample_CCSNe}}
We include host galaxies of all SNe with the following spectroscopic classifications: \texttt{SNIb}/\texttt{c}/\texttt{bc}, \texttt{II}, \texttt{IIb}, \texttt{IIn}. Objects belonging to the SLSN subclasses are treated separately. For the same reasons as with the SNe Ia, we use subsample that was classified with the AAT, resulting in a sample of 47 objects. We compare the DES sample ($\hat{z} = 0.14$) to the subsample of CCSNe from \citet{Kelly2012} (hereafter KK12) that were discovered in untargeted surveys. This results in 117 objects with $\hat{z} = 0.04$. Host galaxy properties were measured from Sloan Digital Sky Survey (SDSS) photometry, using \texttt{P\'EGASE.2} templates for the SED fitting.

\subsubsection{SLSNe \label{subsubsec:sample_SLSNe}}
We include host galaxies of all SLSNe from the sample of \citet{Angus2019} for which a galaxy is detected and a redshift is available. We relax the magnitude-limited selection criterion in order to maintain the sample size. While this means that the sample is not homogeneously selected, we note that the comparison samples have been selected in a similar fashion. We compare the 22 DES SLSN sample to the SLSN samples of PTF (\citealt{Perley2016c}; 32 objects at $\hat{z} = 0.25$) who use a custom SED fitting code using \citet{Bruzual2003} templates, and the combined Pan-STARRS and literature sample of \citet{Lunnan2014}, which we denote PS+ (31 objects, $\hat{z} = 0.64$), who use the \texttt{FAST} SED fitting code \citep{Kriek2009} with \citet{Maraston2005} templates and a Salpeter IMF. We also compare to the DES CCSNe (AAT) sample. 

\subsubsection{RETs \label{subsubsec:sample_RETs}}
Rapidly evolving transients (RETs) are bright events of unknown origin which rise and decline on much faster timescales than classical SNe \citep{Drout2014,Arcavi2018,Pursiainen2018}. We include hosts from the photometrically defined samples of \citealt{Pursiainen2018}, Pursiainen et al., \textit{in prep} for which a host is detected and a redshift is available. The final sample includes 51 objects. The DES RET sample is compared to the Gold and Silver samples from PS1 (\citealt{Drout2014}; 10 objects at $\hat{z} = 0.27$). PS1 stellar masses have been calculated in the same way as the PS+ SLSNe. We also compare DES RETs to the DES CCSNe (AAT) and SNe Ia (AAT) samples.

\section{Host Stellar Mass Distributions} \label{sec:results}
 In this section we construct cumulative distributions of host galaxy stellar masses, and statistically compare the DES samples to those from the literature that have been introduced in Section \ref{subsec:sample}.
The host galaxy magnitudes, redshifts, and derived stellar masses for each DES sample are reported in Tables \ref{tab:Ias}-\ref{tab:RETs}.
\subsection{Probabilistic treatment of mass distributions \label{subsec:res_stats}}

Here we introduce the probabilistic methods used to estimate the true observed stellar mass distributions including upper limits, as well as a Bayesian method to compare the resulting distributions.

The probability density function (PDF) and cumulative density function (CDF) for the host stellar masses of the above-described samples are shown in Figs. \ref{fig:Ia_cum}-\ref{fig:ret_cum}. The CDF represents the cumulative fraction of the total sample of hosts with a stellar mass at or below a given value. The shape of the CDFs of different samples can therefore be used as a comparison, and is often the basis of the `two-sample' tests used to determine if the samples were drawn from the same parent population.

A difficulty arises when for some SNe no host galaxy is detected and only an upper limit is reported. The problem is that the host mass could take any value lower than the upper limit, and thus it is not known at what value the galaxy should be added to the CDF. Typically, to incorporate upper limits in estimating the CDF, astronomers use survival analysis, a technique developed principally for the assessment of the effectiveness of drugs in curing illness. However, most survival analysis focusses on right-censored data - that is, when the survey of patients is conducted, some are still alive, and thus their survival times are unknown. Techniques handling the left-censored data (i.e. upper limits) common to magnitude-limited astronomical surveys are scarce, with authors commonly using historical survival analysis packages such as \texttt{ASURV}\footnote{http://astrostatistics.psu.edu/statcodes/asurv} or the \texttt{Python} package \texttt{lifelines}. Those packages are based around the Kaplan-Meier (KM) estimator \citep{Kaplan1958} of the survival function which approximates the most likely values for the non-detections based on the detected data, and inserts them into the CDF. However, this selection is only performed once, does not incorporate knowledge of the uncertainty on the objects that were detected, and assumes that the non-detections follow the same intrinsic magnitude distribution as the detected data. 

To create our CDFs, we treat both detected points and upper limits as probability distributions. Detections are treated as Gaussians, with a mean and standard deviation corresponding to their detected values and uncertainties. Upper limits are treated as a skewed normal probability distribution, chosen such that the peak of the distribution is aligned with the upper limit minus the mean uncertainty on the detected galaxies. We use a distribution with a skew of $-7$, indicating the distribution is heavily skewed towards the lower end. This way, there is a small but finite probability of the true mass being higher than, but within uncertainty of, the given upper limit. We then 
simulate $10^4$ realisations of the CDF,
each time randomly drawing from the given PDF for each observation. We then take the median, minimum and maximum mass values for each incremental increase in the fraction observed in order to construct the median CDF, and its lower and upper 1$\sigma$ uncertainty.
We find that this method reproduces the CDF given by KM estimation to a good degree, but we are more robust to noisy data having included measurement uncertainties.

To compare host stellar mass distributions, we follow the method described in \citet{Kruschke2012}. We model the PDF of each host sample as skewed-normal distributions. We adopt the \texttt{SciPy} terminology: the skewed-normal distributions are parameterised by their `loc' ($\mu$, analogous to the mean for a distribution with skewness 0), `scale' ($\sigma$, analogous to the standard deviation for a distribution with skewness 0) and `skewness' ($\alpha$, indicating the direction and strength of the skew of the distribution). We begin by assuming as a null hypothesis that both sample distributions are drawn from the same underlying population, and as such choose priors based on the combination of the samples.  For $\mu$ we apply a normal prior, with the hyperparameters set as the mean and double the standard deviation of the combined sample. For $\sigma$ we apply a uniform prior between 0 and the range of masses in the combined sample. For $\alpha$, we apply a weak normal prior centered on -3 with a standard deviation of 5, since we expect host mass distributions to be negatively skewed as the rate of supernovae typically scales with stellar mass. A key advantage over traditional two-sample comparisons such as the Kolmogorov-Smirnov (KS) test is our inclusion of the uncertainty in the likelihood function. By incorporating this uncertainty, we are robustly handling the upper limits and poorly constrained observations so prevalent in observational astronomy. 
We sample from the posterior distribution using the \texttt{pymc3}\footnote{https://docs.pymc.io/} package, using the No-U-Turn Sampler (\texttt{NUTS}; \citealt{Hoffman2014}), which is initialised using \texttt{jitter+adapt\_diag}. We sample with a total of 2000 iterations after a tuning stage of 500 iterations.
For each pair of samples, we then compare the estimates for $\mu$, $\sigma$, and $\alpha$ along with their resulting uncertainties. Since there are some degeneracies between the three parameters, and since they are not physically motivated, we do not calculate a confidence at which the samples are drawn from the same population (i.e. the KS $p$-value). Instead, we quote the probability that each parameter is the same for both samples, and comment on the physical implications. The best fit distributions and parameter confidence intervals can be found in Appendix \ref{app:compare} and \ref{app:fitparams}. The implications of the different host mass distributions for these sample are discussed in Section \ref{subsec:disc_masses}. We stress that these are observed distributions; we do not correct for the numerous selection effects that likely affect each sample.

\begin{figure}
\includegraphics[width=0.5\textwidth]{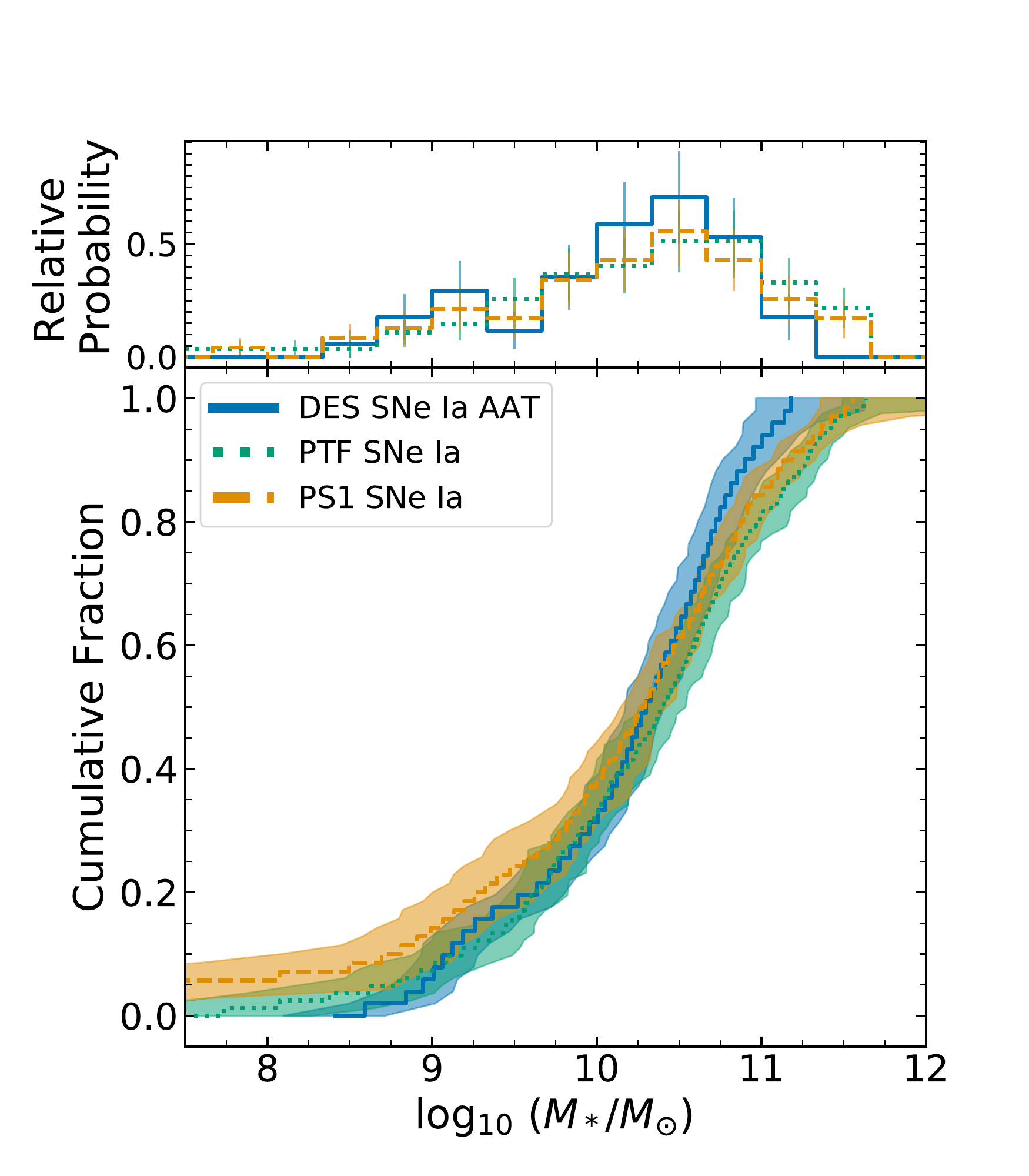}
\caption{As per Fig. \ref{fig:Ia_cum} but restricted to $z<0.2$. The distributions are much more consistent with each other, suggesting that the samples are complete and robust against selection effects at low redshift.
\label{fig:Ia_cum_loz}}
\end{figure}

\subsection{SNe Ia \label{subsec:res_Ia}}

For brevity, in the Results and Discussion sections we refer to the spectroscopic sub-samples defined in Section \ref{subsec:sample} as DES.
Unless otherwise stated, stellar masses are given in units of $\log_{10} \left(M_* / M_{\odot}\right)$. The mean (median) stellar mass of the DES sample is 10.01 (10.15). The average stellar masses from the PS1, PTF, and SNLS samples are similar, but all slightly higher mass than DES: PTF has a mean (median) of 10.25 (10.39); PS1 a mean (median) of 10.07 (10.32); SNLS a mean (median) of 10.08 (10.17). There is a clear difference between the PDFs of DES and the other surveys, with a sharp peak at $\sim10.2$ and a steeper high-mass decline. This manifests in the CDF as a steeper rise and hints at a preference for slightly lower mass galaxies, or that there is a smaller fraction of SNe in higher-mass galaxies. PS1, on the other hand, has a shallower distribution.

The simultaneous fitting of the stellar mass distributions (see Appendix \ref{app:compare} for plots of the posterior samples) shows that the DES and PS1 hosts are clearly distinct. The mean difference in `loc' parameter (which determines the location of the peak of the distribution, but does not correspond directly to the mean, median, or mode), -0.47, is nearly 10 times the standard deviation of the posterior distributions of the differences in loc (0.05). The mean difference in scale is $-0.34\pm0.06$, indicating the DES PDF is significantly narrower. The difference in skewness is less strongly constrained at $-2.3 \pm 1.9$, the stronger DES skewness resulting from the sharper high-mass cut-off. We thus find that the distributions, if assumed to be skewed-normal, are statistically different with the DES (AAT) sample lying at systematically lower $M_*$ due to a stronger avoidance of galaxies with $M_* > 11.5$. 

The DES and PTF Ia host stellar mass distributions are more similar. The difference in their locs is much smaller than for the DES-PS1 comparison, and is only midly significant ($-0.06 \pm 0.05$), as the distributions peak at very similar masses. Similarly, the difference in the scale of the PDFs is small (0.19) but significant ($4\sigma$), with PTF being narrower due to smaller proportion of hosts in the $8-9.5$ range. PTF is thus marginally more strongly skewed, although the significance of this difference is weak when the uncertainty is taken into account. On the whole we suggest these two distributions are likely very similar, while acknowledging minor differences between the DES and PTF distributions around the $1\sigma$ significance.

 The DES and SNLS distributions peak at similar locations, although with the relatively well constrained loc parameter differing by an average of $-0.07 \pm 0.04$ they are different at 1.75$\sigma$ significance with DES located at slightly lower mass. The scales are also different at similar significance ($-0.08 \pm 0.05$), suggesting DES is marginally narrower. The skewness difference is large (mean of -4.9) but uncertain (standard deviation of 2.1), and thus significant at around $2.4\sigma$. As per the PS1 and PTF comparisons, the scale and skewness differences are driven by the DES sample lacking objects above masses of 11.5. The SNLS distribution appears to have a double-peaked nature, and so is not accurately modelled by a single skewed-normal distribution. Thus, the statements of similarity and difference made above are only approximations. However, we note that by treating the distributions with more complex models, we are likely to find even stronger dissimilarities between them. The SN Ia results are discussed in Section \ref{subsubsec:disc_Ia_masses}.
 
 \subsubsection{Comparison to SVA1 stellar masses \label{subsubsec:Ia_vs_sva1}}

Stellar masses for the hosts of SNe Ia in the DES3YR analysis were estimated using photometry from the SVA1 catalogue. In \citetalias{Smith2020} the stellar masses are re-evaluated using SN Deep photometry. The weighted mean stellar mass reduces from $10.64\pm0.06$ to $10.16\pm0.05$. A detailed investigation of the implications of SN Deep catalogue for the SN Ia host galaxy mass step is presented in \citetalias{Smith2020}.

\subsection{CCSNe \label{subsec:res_CC}}

The PDFs and CDFs of CCSNe host masses are shown in Fig. \ref{fig:cc_cum}. 
The CCSNe samples systematically lie at lower mass than the SNe Ia. The DES mean (median) is 9.27 (9.39) while for \citetalias{Kelly2012} this is 9.26 (9.39). The Bayesian fits to their PDFs appear very similar. The mean differences in loc is negligible ($0.00 \pm 0.09$), while DES is slightly narrower at $-0.24 \pm 0.09$ ($\sim 2.5\sigma$). There is a hint of a difference in skewness: while DES is strongly skewed ($-7.00 \pm 3.58$), \citetalias{Kelly2012} is less so ($-4.29 \pm 2.07$). However the broad posterior distributions for the skewness allows for the difference to be consistent with zero ($-2.71 \pm 4.15$). As with the SNLS SNe Ia, the DES CCSNe mass distribution has a strong double-peaked nature, meaning the PDF is not well approximated by a single skewed-normal distribution. As a result, although the fits provide a good match between the two host distributions - we do not statistically determine them to be different - we exercise caution and suggest that the double-peaked nature may be due to strong selection effects, which we discuss in Section \ref{subsubsec:disc_cc_masses}.

\subsection{SLSNe \label{subsec:res_SLSNe}}

The host galaxies of SLSNe are on average lower in stellar mass than those of CCSNe. The means (medians) of the DES, PTF and PS+ samples being 8.67 (8.58), 8.75 (8.70), and 8.35 (8.19) respectively.

 The skewed-normal fits to the SLSN samples are relatively unconstrained due to their small sizes (22, 32 and 31 objects respectively, many of which are upper limits), particularly for the DES sample.  However there are some notable differences. The SLSN sample CDFs all appear to follow a similar shape to each other, rising steeply rising at lower mass, while the CCSN sample is shallower and peaks at higher mass.

We are unable to fit a well-constrained skewed-normal distribution to the DES sample, and thus do not make claims on its statistical similarity to the comparison samples.

\subsection{RETs \label{subsec:res_RETs}}

The mean (median) RET host mass is 9.59 (9.41) for DES and 9.20 (9.04) for PS1. This is significantly larger than the SLSN samples, and consistent with the CCSNe. 
Rapid transient hosts do not appear to follow the CDFs of SNe Ia, CCSNe, or SLSNe. There is a plateau in the DES RETs CDF around $\log_{10} \left(M_* / M_{\odot}\right) = 10$ which bears resemblance to the PTF SLSNe CDF. However, the RET hosts CDF rises again, with around 20\% of the galaxies lying above 10.5. 

A simple skewed-normal distribution does not fit the PS1 RET PDF particularly well due to the sample size of only 10 objects. Inspecting the CDF shows that the DES and PS1 RET host mass distributions are consistent within errors, particularly at the high-mass end.

The DES RET PDF is most similar to the CCSNe. There are still strong differences: the RETs are located at lower mass (difference in loc $-0.70 \pm 0.29$) and have a much narrower distribution (difference in scale $-0.66 \pm 0.10$). The RET distribution somewhat negatively skewed, with an average alpha of $-1.78$, but this is highly uncertain ($\pm 2.88$) due to a small but significant tail of the posterior distribution extending to very strong skewnesses, effectively cutting off the PDF at a mass limit of 10. The differences between the DES RETs and DES SNe Ia are even stronger, since the Ia distribution is shifted to even higher mass than the CCSNe - the lower mass loc in DES RETs ($-1.32 \pm 0.27$) is significant at $4.8\sigma$; the narrower DES RET distribution ($-0.75 \pm 0.08$) at $9.4\sigma$. While they occur on average in lower mass hosts than SNe Ia and CCSNe, the DES RET hosts appear to be significantly higher in mass than the DES SLSNe, although given that the skewed-normal fit to the DES SLSN PDF is largely unconstrained we do not claim that they are statistically different.
We thus conclude that the DES RET host mass sample is significantly different to the DES samples of CCSNe and SNe Ia, and visually different to SLSNe.

\begin{figure}
\includegraphics[width=0.5\textwidth]{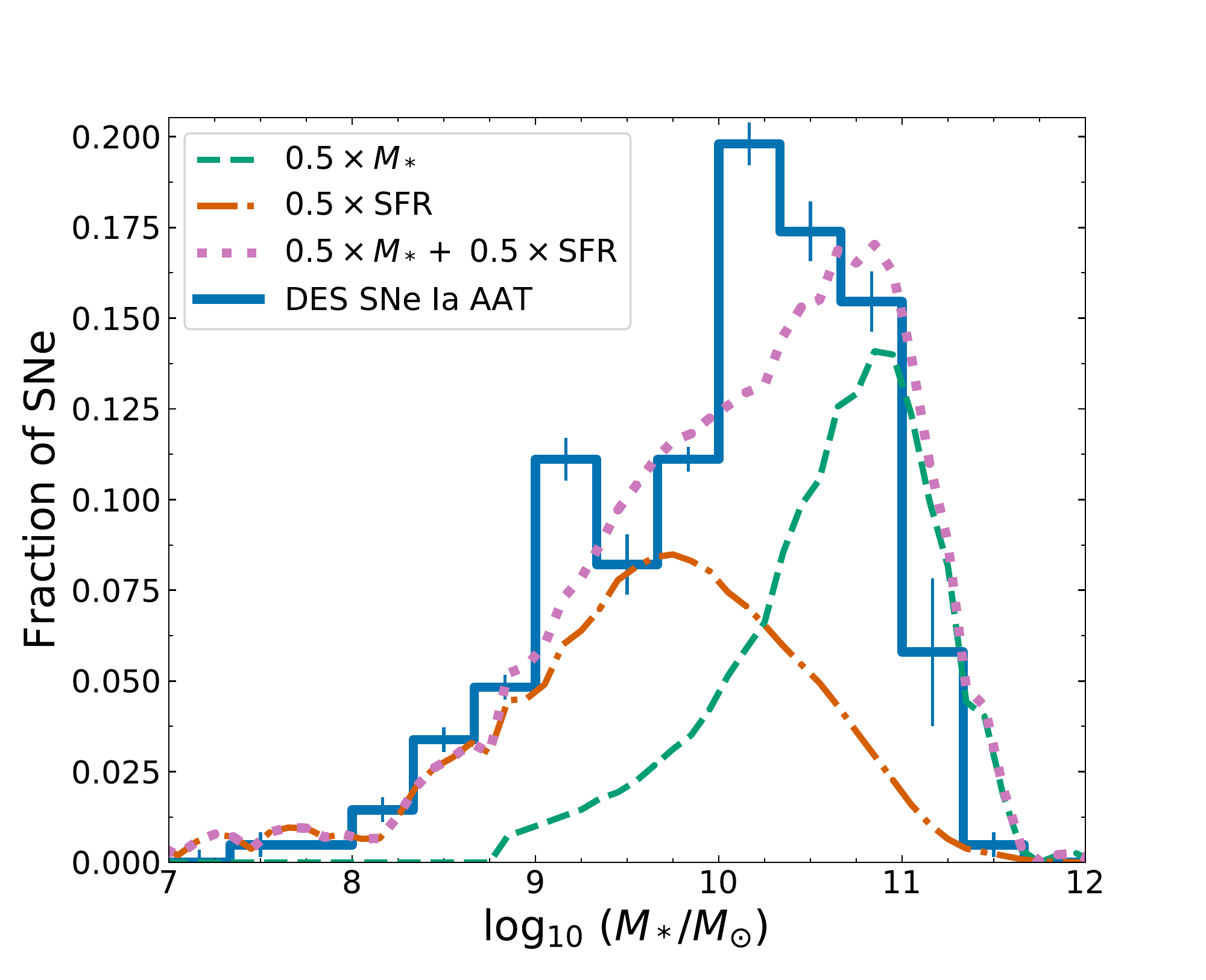}

\caption{The DES SNe Ia (AAT) host stellar mass distribution as shown in Fig. \ref{fig:Ia_cum} (blue, solid line) compared to the relative distribution of stellar mass ($M_*$, based on the $K$-band survey of \citealt{Beare2019}) and the distribution of star formation rate (SFR) in terms of the stellar mass of star-forming galaxies} from SDSS \citep{Abazajian2009}. $M_*$ and SFR have been scaled by 0.5 such that their sum is 1. 
\label{fig:Ia_mass_pred}
\end{figure}

\begin{figure}
\includegraphics[width=0.5\textwidth]{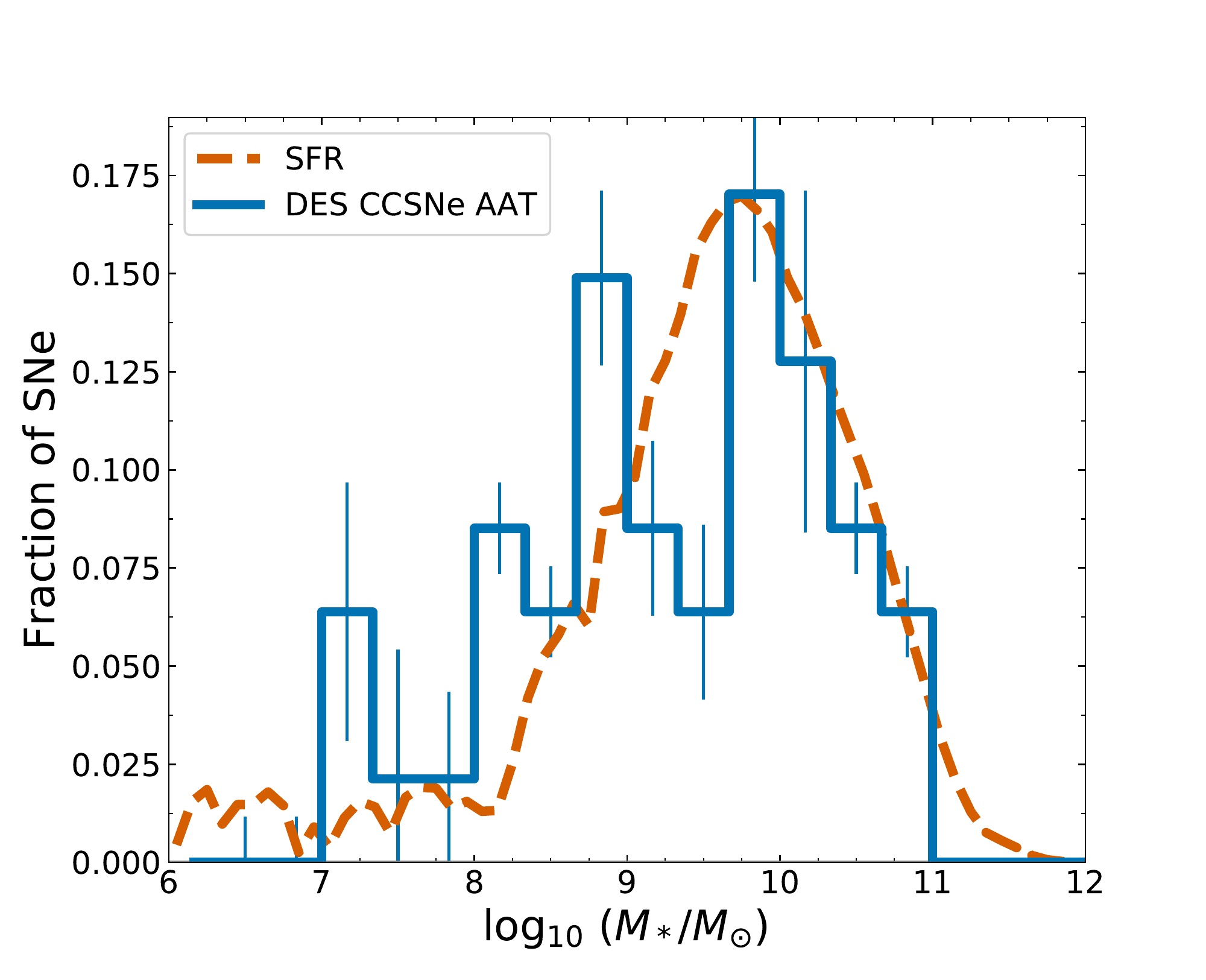}

\caption{The DES CCSNe (AAT) host stellar mass distribution as shown in Fig. \ref{fig:cc_cum}, compared to the relative mass distribution of star formation from SDSS. The CCSNe appear to trace the mass distribution of the SFR. } 
\label{fig:cc_mass_pred}
\end{figure}

\begin{figure}
    \includegraphics[width=0.5\textwidth]{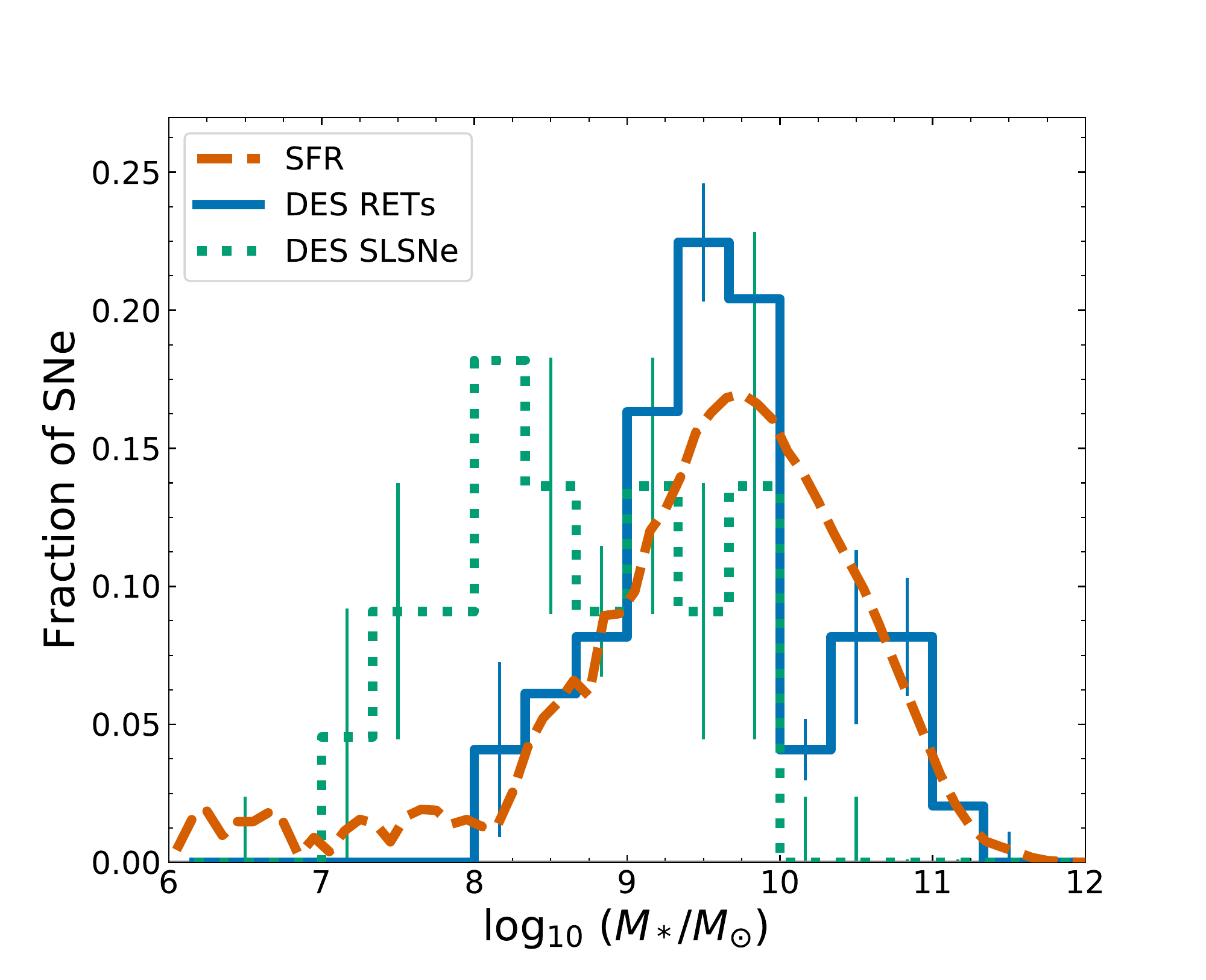}

\caption{The DES RET and SLSN host stellar mass distributions as shown in Figs. \ref{fig:ret_cum},\ref{fig:slsn_cum}, compared to the relative mass distribution of star formation from SDSS. The SLSNe distribution is distinct from the SFR, but the distinction for RETs is not so clear. } 
\label{fig:ret_slsn_mass_pred}
\end{figure}

\section{Discussion} \label{sec:discussion}
 \subsection{Coadds \label{subsec:disc_coadds}}
We have created the a set of deep images in order to obtain precise and accurate measurements of DES supernova host galaxies, a subset of which are presented in a catalogue here. We have optimised the coadds almost entirely for depth, without focussing much attention on the resulting PSF, nor on the removal of the faintest of artefacts. For that reason, these coadds are not optimal for studies such as weak lensing or those which require accurate modelling of galaxy morphology and light profiles. For those studies, we refer the reader to DES Collaboration 2019 (\textit{in prep}).
However, Figs. \ref{fig:stamp_eg}, \ref{fig:magdiff_combined}, \ref{fig:magerrdiff_combined},  and \ref{fig:compare_mag_dists} are clear evidence for success of the depth-optimised coadds - we detect fainter hosts, thus improving the fraction of correctly assigned hosts. We also reduce the error compared to the shallower SV catalogues as well the deep catalogue that has not used a depth optimisation.

To our knowledge, we thus present the highest volume host galaxy catalog for an untargeted transient survey, which will allow a thorough exploration of host magnitude and mass distributions, misidentification fraction and spectroscopic/photometric redshift efficiency, to be presented in future work along with the DES-SN 5YR cosmological analysis.

A further advantage of our work is the ease with which the selection of input images can be optimised for desired characteristics in the resulting coadd. We refer the reader to \citet{Kelsey2020}, who optimise the stacks for resultant seeing. 

 \subsection{Host Stellar Masses \label{subsec:disc_masses}}
 \subsubsection{SNe Ia \label{subsubsec:disc_Ia_masses}}
 
 The results presented in Section \ref{subsec:res_Ia} show the host galaxy stellar masses of the DES SNe Ia (AAT) sample to be significantly different from that of the PS1 and SNLS samples, consistent with the conclusion derived by \citetalias{Smith2020} using the KS test. The shape is more similar to that of the PTF sample. 
 
DES SNe Ia (AAT) hosts lie, on the whole, at lower mass, with a tendency to avoid galaxies above $\log_{10}\left(M_*/M_{\sun}\right) \sim 11.5$. One reason for this could be the selection criteria for the DES sample. All SNe included in this sample were classified using the AAT, a 4m telescope using a fibre-fed spectrograph. It is likely that the spectra of some of the SNe that occurred in higher-mass galaxies are dominated by relatively bright host galaxy continuum, hindering a classification. The 5-year, photometrically selected sample should be devoid of these effects. On the other hand, the lack of inclusion of hostless objects and those missing a host-galaxy redshift in this sample is likely to have added a bias in the other direction: that is, apparently hostless SNe must have exploded in a region of low stellar mass for no host to have been detected. A future analysis will develop on the survival analysis developed here in order to include objects with no redshift, while the effects of the host galaxy selection function will be explored in M\"oller et al. \textit{in prep}. 

The differences between the distributions could be caused in part by the selection function at different redshifts. To test this, we restricted the redshifts to $z<0.2$. The resulting CDF is shown in Fig. \ref{fig:Ia_cum_loz} and shows that the DES, PS1 and PTF samples are broadly consistent at low-$z$. The divergence in the full CDFs is therefore attributable to the high-$z$ selection function, which is a combination of the SN detection and classification efficiencies (which are dependent on host galaxy mass), as well as the host galaxy detection completeness and redshift measurement completeness.

SNe Ia are thought to arise in various types of stellar population: the young, `prompt' SNe Ia, associated with young stellar populations and thus in star-forming galaxies; and the older, `delayed' population, which occur in old stellar populations \citep[e.g.][]{Scannapieco2005,Sullivan2006,Sullivan2010,Childress2014}, and thus are found in both star-forming and passive galaxies. The shape of the host galaxy mass distribution at a given redshift is thus expected to be shaped by a combination of the mass-distribution of the star-formation rate (SFR) and the raw stellar-mass distribution, and the relative contributions of these two distributions can help constrain the distribution of SN Ia progenitors among stellar populations. As per \citet{Childress2013,Childress2013a}, in Fig. \ref{fig:Ia_mass_pred} we show the DES SNe Ia host stellar mass PDF compared to a sum of the stellar mass distribution of SFR from SDSS Data Release 7 \citep{Abazajian2009}, and to the raw stellar mass distribution as measured by near-infrared photometry  \citep{Beare2019}. We cut the samples to the appropriate redshift range to match the DES SN Ia sample. We scale the SFR and raw $M_*$ distributions by half, such that when summed their distributions approximate the total SN Ia host mass distrubtion. The SN Ia host mass distribution resembles this combination well. The location of the sub-peak at $\log_{10} \left(M_*/M_{\sun}\right) \sim 9$ is in the regime where the SN Ia rate is dominated by star formation rather than stellar mass, the so-called `prompt' population. We do not perform a fit of the host mass function to the SFR plus stellar mass combination, but an analysis with the full five year, photometrically selected DES-SN data set will be presented in future work.

\subsubsection{CCSNe \label{subsubsec:disc_cc_masses}}
The DES and \citetalias{Kelly2012} CCSNe host mass distributions are not statistically different from one another, based on the Bayesian fitting of a skewed-normal distribution. However, visual inspection shows the DES sample to be double-peaked, with an apparent lack of hosts with masses of $\log_{10} \left(M_*/M_{\sun}\right) \sim 9.5-10$. This may be a by-product of the selection function, as CCSNe are in general fainter than SNe Ia, as well as having fewer strongly distinguishing features, such that they are even more difficult to classify when the spectrum is dominated by a massive host galaxy. However this would be more apparent as a high-mass cut-off rather than a double peak. We anticipate that a larger, photometrically selected DES CCSNe sample with smaller selection effects will be able to clarify the nature of this distribution, and will present that analysis in a future work.  

CCSNe occur in regions of ongoing star formation, and thus the stellar masses of their host galaxies are expected to trace the mass distribution of star-forming galaxies, without the higher-mass contribution from the raw stellar mass that is apparent in the SN Ia host distribution. In Fig \ref{fig:cc_mass_pred} we plot the CCSN host mass distribution compared to the mass distribution of star formation. The location of the peaks is roughly consistent around $\log_{10}\left(M_*/M_{\sun}\right) = 10$, while the maximum CCSN host mass is consistent with the upper galaxy mass at which there is significant star formation, suggesting there is no obvious other factor (such as metallicity) inhibiting the production of CCSNe, which we might see if we had split the samples up into sub-types.

\subsubsection{SLSNe
\label{subsubsec:disc_slsn_masses}}
SLSNe show a strong preference for low mass, low metallicity, highly starforming host galaxies \citep[e.g.][]{Lunnan2014,Leloudas2015,Chen2016a,Angus2016}. We have shown in Fig. \ref{fig:slsn_cum} that for stellar mass, the same is true for the DES sample. This result is reinforced by Fig. \ref{fig:ret_slsn_mass_pred}, where the SLSN host mass PDF has a cut-off at much lower masses than the SFR distribution. The host masses are consistent with the PS+ and PTF samples, despite the DES objects themselves showing strong lighturve and spectral diversity, extending to fainter luminosities than the PS+ sample in particular, and covering a broader redshift range \citep{Angus2019}. Recently, a handful of SLSNe have been discovered in high-mass hosts \citep{Chen2017b,Izzo2017}; future large, volume limited complete samples from the Zwicky Transient Facility (ZTF) and the Large Synoptic Sky Survey (LSST) will be able to determine whether these high mass hosts are statistical anomalies, or whether previous and current studies were biased strongly by selection criteria.

\subsubsection{RETs
\label{subsubsec:disc_ret_masses}}
The DES sample of RETs is the largest to date by an order of magnitude. The hosts in the sample analysed here appear to have a narrow stellar mass distribution with a significant high-mass tail, unlike either SNe Ia, CCSNe, or SLSNe. They do not appear to trace the mass distribution of SFR (Fig. \ref{fig:ret_slsn_mass_pred}), although they follow it more closely than SLSNe, particularly at lower masses. It is unclear how the atypical shape of the distribution, in particular the lack of hosts at $\log_{10}\left(M_*/M_{\sun}\right) = 10$, could be caused by a selection effect. Typically selection biases have a smooth effect on the PDF, rather than the strong cut-off seen in the RET sample. There is a possibility that the effect is physical, such as the RETs belonging to more than one population of transients. In this scenario, a dominant population of transients could follow the SFR at low masses, but be subject to a metallicity threshold, while a sub-dominant population trace instead stellar mass and lead to the high-mass tail. This is certainly plausible, as the lightcurve shapes and luminosities of the RETs show strong diversity \citep{Pursiainen2018}. A future investigation into correlations between RET and host galaxy properties will provide further insight into this possibility.

\section{Conclusions}  \label{sec:conclusion}

In this work, we have created a set of coadded optical images in the fields of the DES-SN programme, optimized for their ultimate depth. Simultaneously, we have created a framework with which it is possible to stack while optimizing for diagnostics of choice, such as seeing. With photometry reaching depths in excess of 27th magnitude, these coadds provide the basis for the ongoing analysis of the SN Ia host galaxy mass step in the DES3YR cosmological dataset (\citetalias{Smith2020}; \citealt{Kelsey2020}), and lay the foundation for the full, photometrically selected analysis, currently in progress.
Along with the direct use in the cosmological fit, these coadds and the derived galaxy catalogues provide room for exploration into further correlations between SN properties and their host galaxies.

The secondary outcome of this work is the comparison between the host masses of various samples of DES supernovae, and their corresponding samples from other surveys. By employing Monte Carlo-based techniques, we have allowed for the inclusion of uncertainties and limits in the construction of cumulative distributions. We further the analysis by creating a probabilistic framework with which to then compare distributions in a non-parametric manner, while taking into account the uncertainties derived in their construction.
We use this framework to infer that the DES type Ia supernova host galaxy stellar mass distribution is different to that from PS1 and SNLS, and more similar in shape to the PTF sample. The core-collapse supernova host galaxy sample is statistically similar to a low redshift compilation. We are unable to determine the degree to which these similarities and differences are inherent to the underlying host galaxy populations, or to the various selection biases associated with spectroscopically selected samples. The DES superluminous supernova hosts are similar in stellar mass to the PS+ and PTF samples, while the rapidly evolving transient hosts are somewhat different to all of the other samples analysed, hinting at the possibility of multiple underlying transient and/or host galaxy populations.

Adding significant ancillary benefits to the project, deep colour images created from the DES-SN coadds have been used at public engagement events across the UK showcasing the importance and impact of surveys such as DES.

\section*{Acknowledgements}

We acknowledge support from STFC grant ST/R000506/1 and EU/FP7-ERC grant 615929. L.G. was funded by the European Union's Horizon 2020 research and innovation programme under the Marie Sk\l{}odowska-Curie grant agreement No. 839090.

This paper makes use of observations taken using the Anglo-Australian Telescope under programs ATAC A/2013B/12 and NOAO 2013B-0317.

This research used resources of the National Energy Research Scientific Computing Center (NERSC), a U.S. Department of Energy Office of Science User Facility operated under Contract No. DE-AC02-05CH11231.

Funding for the DES Projects has been provided by the U.S. Department of Energy, the U.S. National Science Foundation, the Ministry of Science and Education of Spain, 
the Science and Technology Facilities Council of the United Kingdom, the Higher Education Funding Council for England, the National Center for Supercomputing 
Applications at the University of Illinois at Urbana-Champaign, the Kavli Institute of Cosmological Physics at the University of Chicago, 
the Center for Cosmology and Astro-Particle Physics at the Ohio State University,
the Mitchell Institute for Fundamental Physics and Astronomy at Texas A\&M University, Financiadora de Estudos e Projetos, 
Funda{\c c}{\~a}o Carlos Chagas Filho de Amparo {\`a} Pesquisa do Estado do Rio de Janeiro, Conselho Nacional de Desenvolvimento Cient{\'i}fico e Tecnol{\'o}gico and 
the Minist{\'e}rio da Ci{\^e}ncia, Tecnologia e Inova{\c c}{\~a}o, the Deutsche Forschungsgemeinschaft and the Collaborating Institutions in the Dark Energy Survey. 

The Collaborating Institutions are Argonne National Laboratory, the University of California at Santa Cruz, the University of Cambridge, Centro de Investigaciones Energ{\'e}ticas, 
Medioambientales y Tecnol{\'o}gicas-Madrid, the University of Chicago, University College London, the DES-Brazil Consortium, the University of Edinburgh, 
the Eidgen{\"o}ssische Technische Hochschule (ETH) Z{\"u}rich, 
Fermi National Accelerator Laboratory, the University of Illinois at Urbana-Champaign, the Institut de Ci{\`e}ncies de l'Espai (IEEC/CSIC), 
the Institut de F{\'i}sica d'Altes Energies, Lawrence Berkeley National Laboratory, the Ludwig-Maximilians Universit{\"a}t M{\"u}nchen and the associated Excellence Cluster Universe, 
the University of Michigan, the National Optical Astronomy Observatory, the University of Nottingham, The Ohio State University, the University of Pennsylvania, the University of Portsmouth, 
SLAC National Accelerator Laboratory, Stanford University, the University of Sussex, Texas A\&M University, and the OzDES Membership Consortium.

Based in part on observations at Cerro Tololo Inter-American Observatory, National Optical Astronomy Observatory, which is operated by the Association of 
Universities for Research in Astronomy (AURA) under a cooperative agreement with the National Science Foundation.

The DES data management system is supported by the National Science Foundation under Grant Numbers AST-1138766 and AST-1536171.
The DES participants from Spanish institutions are partially supported by MINECO under grants AYA2015-71825, ESP2015-66861, FPA2015-68048, SEV-2016-0588, SEV-2016-0597, and MDM-2015-0509, 
some of which include ERDF funds from the European Union. IFAE is partially funded by the CERCA program of the Generalitat de Catalunya.
Research leading to these results has received funding from the European Research
Council under the European Union's Seventh Framework Program (FP7/2007-2013) including ERC grant agreements 240672, 291329, and 306478.
We  acknowledge support from the Brazilian Instituto Nacional de Ci\^encia
e Tecnologia (INCT) e-Universe (CNPq grant 465376/2014-2).

This manuscript has been authored by Fermi Research Alliance, LLC under Contract No. DE-AC02-07CH11359 with the U.S. Department of Energy, Office of Science, Office of High Energy Physics.




\bibliographystyle{mnras}
\bibliography{PhilMendeley} 


\appendix
\suppressfloats[t]
\restylefloat{table}
\onecolumn
\section{Coadds Summary}

\begin{table}[h]
\caption{Overview of the deep coadds. $N_{\textrm{exp}}$ is the number of single exposures in the coadd; $t_{\textrm{exp;tot}}$ is the total exposure time in hours; $m_{\textrm{lim}}$ is the limiting magnitude determined from the sky background. The full table is available online.} 
\label{tab:coadds}
\begin{tabular}{lllllllr}
\toprule
 Field & Band & MY & $\tau_{\textrm{Cut}}$ & PSF Cut & $N_{\textrm{exp}}$ & $t_{\textrm{exp;tot}}$ &  $m_{\textrm{lim}}$ \\
\midrule

\midrule

\bottomrule

 SN-E1 &    g &  1 &                      0.26 &     2.4 &                 70 &                    3.4 &               25.44 \\
 SN-E1 &    g &  2 &                      0.26 &     2.4 &                 48 &                   2.33 &               24.55 \\
 SN-E1 &    g &  3 &                      0.26 &     2.4 &                 54 &                   2.62 &               25.46 \\
 SN-E1 &    g &  4 &                      0.26 &     2.4 &                 52 &                   2.53 &               25.41 \\
 SN-E1 &    g &  5 &                      0.26 &     2.4 &                 60 &                   2.92 &               25.44 \\
 \midrule
 SN-E1 &    r &  1 &                       0.2 &     2.2 &                 90 &                   3.75 &               25.51 \\
 SN-E1 &    r &  2 &                       0.2 &     2.2 &                 69 &                   2.88 &               25.05 \\
 SN-E1 &    r &  3 &                       0.2 &     2.2 &                 69 &                   2.88 &               25.52 \\
 SN-E1 &    r &  4 &                       0.2 &     2.2 &                 63 &                   2.62 &               25.42 \\
 SN-E1 &    r &  5 &                       0.2 &     2.2 &                 73 &                   3.04 &               25.41 \\
 \midrule
 SN-E1 &    i &  1 &                       0.2 &     2.2 &                 95 &                   5.28 &               25.43 \\
 SN-E1 &    i &  2 &                       0.2 &     2.2 &                 74 &                   4.11 &               25.24 \\
 SN-E1 &    i &  3 &                       0.2 &     2.2 &                 72 &                    4.0 &               25.40 \\
 SN-E1 &    i &  4 &                       0.2 &     2.2 &                 68 &                   3.78 &               25.36 \\
 SN-E1 &    i &  5 &                       0.2 &     2.2 &                 75 &                   4.17 &               25.41 \\
 \midrule
 SN-E1 &    z &  1 &                       0.3 &     2.2 &                171 &                    9.5 &               25.23 \\
 SN-E1 &    z &  2 &                       0.3 &     2.2 &                132 &                   7.33 &               25.05 \\
 SN-E1 &    z &  3 &                       0.3 &     2.2 &                130 &                   7.22 &               25.24 \\
 SN-E1 &    z &  4 &                       0.3 &     2.2 &                125 &                   6.94 &               25.17 \\
 SN-E1 &    z &  5 &                       0.3 &     2.2 &                138 &                   7.67 &               25.22 \\
 \end{tabular}
\end{table}


\section{Host Galaxy Data}

\begin{table}[h]
\caption{Host galaxy properties for the DES SN Ia sample. Magnitudes are given in the observer frame and are not corrected for host galaxy extinction. The full table can be accessed in the online version of this manuscript.}
\label{tab:Ias}
\begin{tabular}{lrrrrrrrrrrrl}
\toprule
       Name &      z &   $g$ &  $g$\_err &   $r$ &  $r$\_err &   $i$ &  $i$\_err &   $z$ &  $z$\_err &     M &  M\_err \\
\midrule
  DES13E1ao &   0.17 & 22.84 &      0.02 & 22.32 &      0.01 & 22.18 &      0.02 & 22.08 &      0.02 &  8.41 &   0.02  \\
 DES13C3dgs &   0.35 & 21.80 &      0.00 & 21.04 &      0.00 & 20.81 &      0.00 & 20.59 &      0.00 &  9.57 &   0.01  \\
  DES13S1qv &   0.18 & 22.17 &      0.01 & 21.60 &      0.01 & 21.37 &      0.01 & 21.25 &      0.01 &  8.79 &   0.05  \\
 DES13C1juw &   0.20 & 22.18 &      0.01 & 21.13 &      0.00 & 20.68 &      0.01 & 20.53 &      0.01 &  9.43 &   0.02  \\
 DES13X3woy &   0.32 & 20.23 &      0.00 & 18.75 &      0.00 & 18.16 &      0.00 & 17.98 &      0.00 & 11.16 &   0.02  \\

\bottomrule
\end{tabular}
\end{table}

\begin{table}[h]
\caption{Host galaxy properties for the DES CCSN sample. Magnitudes are given in the observer frame and are not corrected for host galaxy extinction. The full table can be accessed in the online version of this manuscript.}
\label{tab:CCs}
\begin{tabular}{lrrrrrrrrrrrl}
\toprule
       Name &      z &   $g$ &  $g$\_err &   $r$ &  $r$\_err &   $i$ &  $i$\_err &   $z$ &  $z$\_err &     M &  M\_err\\
\midrule
 DES13C3ui & 0.07 & 20.80 &      0.00 & 20.54 &      0.00 & 20.42 &      0.00 & 20.35 &      0.01 &  8.22 &    0.01 \\
 DES13C1feu & 0.06 & 16.30 &      0.00 & 15.61 &      0.00 & 15.42 &      0.00 & 15.24 &      0.00 & 10.29 &    0.01\\
 DES13X3fca & 0.10 & 17.70 &      0.00 & 17.00 &      0.00 & 16.76 &      0.00 & 16.74 &      0.00 & 10.07 &    0.02 \\
  DES15C3bj & 0.29 & 20.80 &      0.00 & 20.07 &      0.00 & 19.72 &      0.00 & 19.54 &      0.00 &  9.98 &    0.01  \\
  DES15S1by & 0.13 & 20.18 &      0.00 & 19.72 &      0.00 & 19.50 &      0.00 & 19.36 &      0.00 &  9.23 &    0.05  \\

\bottomrule
\end{tabular}
\end{table}

\begin{table}[h]
\caption{Host galaxy properties for the DES SLSN sample. Magnitudes are given in the observer frame and are not corrected for host galaxy extinction. The full table can be accessed in the online version of this manuscript.}
\label{tab:SLSNe}
\begin{tabular}{lrrrrrrrrrrrl}
\toprule
       Name &      z &   $g$ &  $g$\_err &   $r$ &  $r$\_err &   $i$ &  $i$\_err &   $z$ &  $z$\_err &     M &  M\_err\\
\midrule
 DES13S2cmm & 0.66 & 24.12 &      0.06 & 23.44 &      0.04 & 22.99 &      0.03 & 23.20 &      0.06 & 8.86 &    0.09 \\
  DES15S2nr & 0.22 & 23.82 &      0.05 & 23.51 &      0.05 & 23.26 &      0.05 & 23.01 &      0.06 & 8.05 &    0.06 \\
 DES14E2slp & 0.51 & 23.63 &      0.04 & 22.62 &      0.02 & 22.32 &      0.02 & 22.06 &      0.03 & 9.36 &    0.04 \\
 DES15S1nog & 0.57 & 23.43 &      0.03 & 22.67 &      0.02 & 22.33 &      0.02 & 22.21 &      0.02 & 9.24 &    0.02 \\
 DES16C3dmp & 0.57 & 22.25 &      0.01 & 21.59 &      0.00 & 21.31 &      0.00 & 21.26 &      0.01 & 9.56 &    0.01 \\

\bottomrule
\end{tabular}
\end{table}

\begin{table}[h]
\caption{Host galaxy properties for the DES RET sample. Magnitudes are given in the observer frame and are not corrected for host galaxy extinction. The full table can be accessed in the online version of this manuscript.}
\label{tab:RETs}
\begin{tabular}{lrrrrrrrrrrrl}
\toprule
       Name &      z &   $g$ &  $g$\_err &   $r$ &  $r$\_err &   $i$ &  $i$\_err &   $z$ &  $z$\_err &     M &  M\_err \\
\midrule
 DES13X3gms & 0.65 & 23.62 &      0.02 & 23.02 &      0.01 & 22.61 &      0.02 & 22.61 &      0.02 &  9.13 &    0.03 \\
  DES13C1tgd & 0.20 & 21.28 &      0.00 & 20.31 &      0.00 & 19.75 &      0.00 & 19.59 &      0.00 &  9.90 &    0.01 \\
  DES13S2wxf & 0.57 & 22.16 &      0.01 & 21.31 &      0.01 & 20.98 &      0.01 & 20.92 &      0.01 &  9.83 &    0.02 \\
  DES13X1hav & 0.58 & 24.38 &      0.09 & 23.63 &      0.07 & 23.24 &      0.04 & 23.13 &      0.05 &  8.95 &    0.07 \\
  DES13X3nyg & 0.71 & 23.84 &      0.03 & 23.39 &      0.02 & 22.96 &      0.02 & 22.92 &      0.03 &  9.06 &    0.03 \\

\bottomrule
\end{tabular}
\end{table}
\begin{multicols}{2}
\section{Stellar Mass Fits}\label{app:compare}
Here we show the Bayesian fits that were used to compare between host stellar mass distributions as outlined in Section \ref{subsec:res_stats}. The solid lines are taken from each of the MCMC samples.

\begin{wrapfigure}{1}{1.\linewidth}
    \includegraphics[width=\hsize]{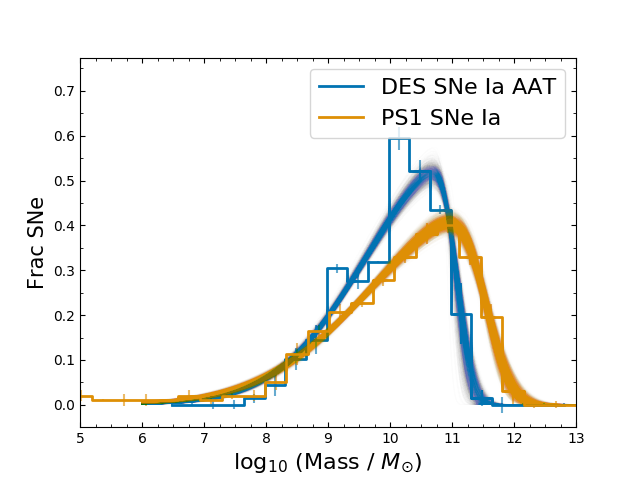}
    \caption{DES SNe Ia compared to PS1 SNe Ia}
    \label{fig:des_Ia_vs_PS1_Ia}
 \end{wrapfigure}
 \vspace{-0.2in}
 \begin{wrapfigure}{1}{1.\linewidth}
    \includegraphics[width=\hsize]{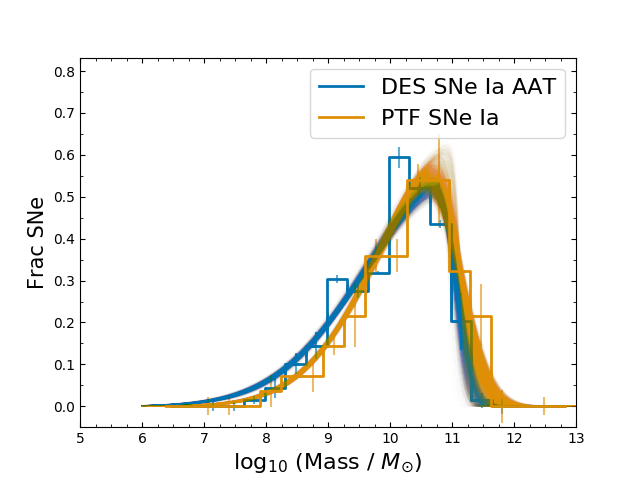}
    \caption{DES SNe Ia compared to PTF SNe Ia}
    \label{fig:des_Ia_vs_PTF_Ia}
 \end{wrapfigure}
 \vspace{-0.3in}
 \begin{wrapfigure}{1}{1.\linewidth}
    \includegraphics[width=\hsize]{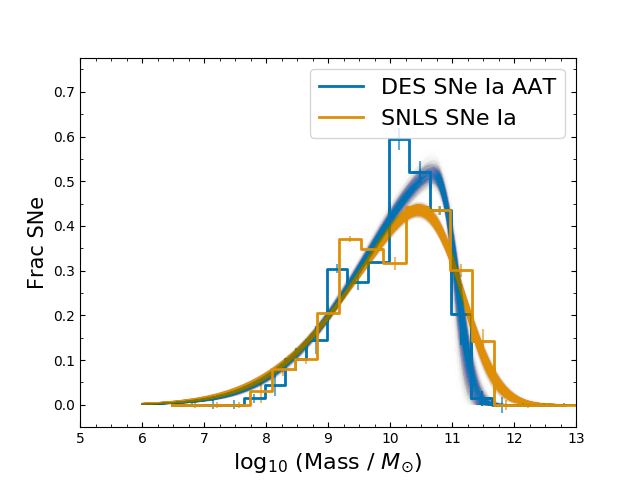}
    \caption{DES SNe Ia compared to SNLS SNe Ia}
    \label{fig:des_Ia_vs_SNLS_Ia}
  \end{wrapfigure}
  \vspace{-0.3in}
 \begin{wrapfigure}{1}{1.\linewidth}
    \includegraphics[width=\hsize]{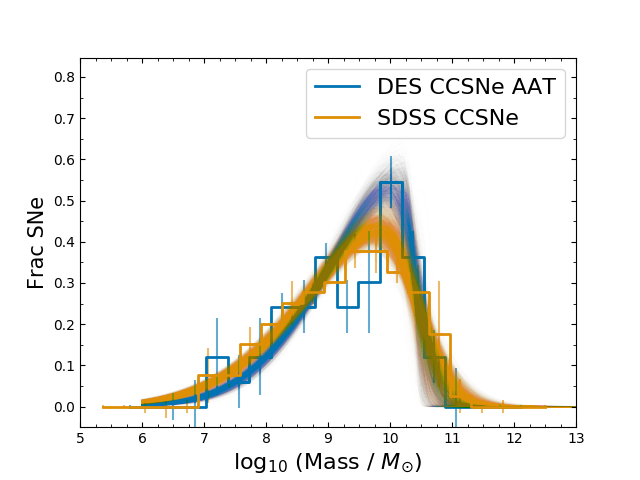}
    \caption{DES CCSNe compared to PTF CCSNe }
    \label{fig:des_CCSN_vs_PTF_CCSN}
 \end{wrapfigure}
 \vspace{-0.3in}
 \begin{wrapfigure}{1}{1.\linewidth}
    \includegraphics[width=\hsize]{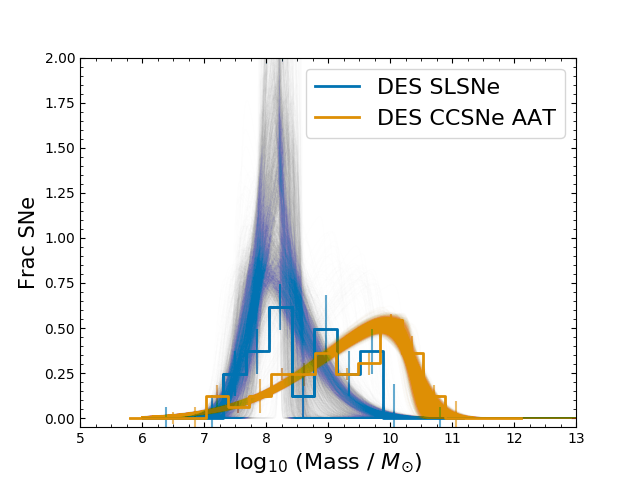}
    \caption{DES SLSNe compared to DES CCSNe}
    \label{fig:des_SLSN_vs_des_CCSN}
  \end{wrapfigure}
  \vspace{-0.3in}
 \begin{wrapfigure}{1}{1.\linewidth}
    \includegraphics[width=\hsize]{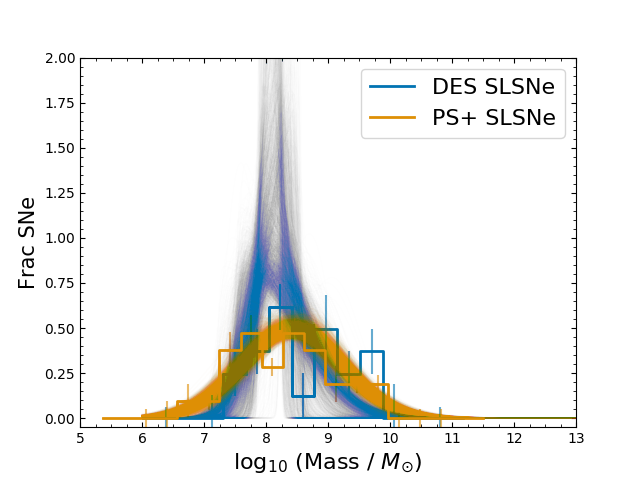}
    \caption{DES SLSNe compared to PS+ SLSNe}
    \label{fig:des_SLSNe_vs_PS1_SLSNe}
 \end{wrapfigure}
 \vspace{-0.3in}
 \begin{wrapfigure}{1}{1.\linewidth}
    \includegraphics[width=\hsize]{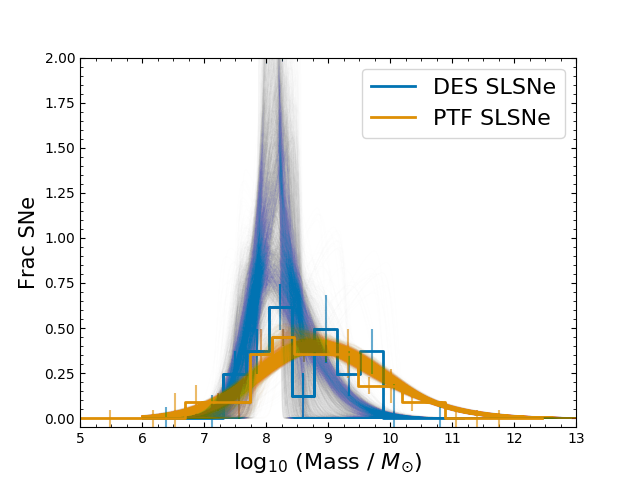}
    \caption{DES SLSNe compared to PTF SLSNe}
    \label{fig:des_SLSNe_vs_PTF_SLSNe}
 \end{wrapfigure}
 \vspace{-0.3in}
 \begin{wrapfigure}{1}{1.\linewidth}
\includegraphics[width=\hsize]{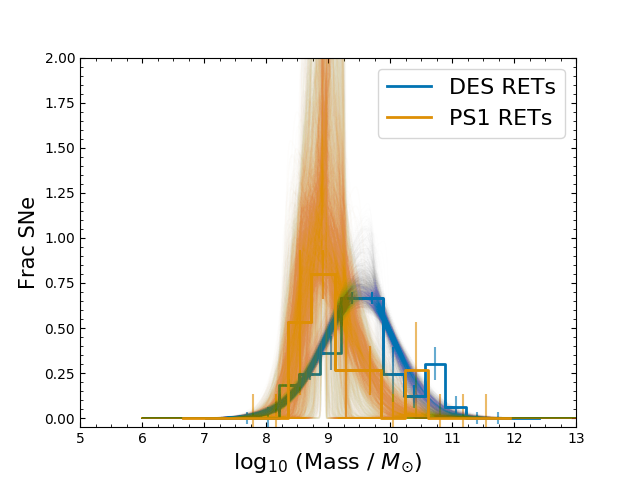}
    \caption{DES RETs compared to PS1 RETs classified}
    \label{fig:des_RETs_vs_PS1_RETs}
 \end{wrapfigure}
 \vspace{-0.3in}
 \begin{wrapfigure}{1}{1.\linewidth}
    \includegraphics[width=\hsize]{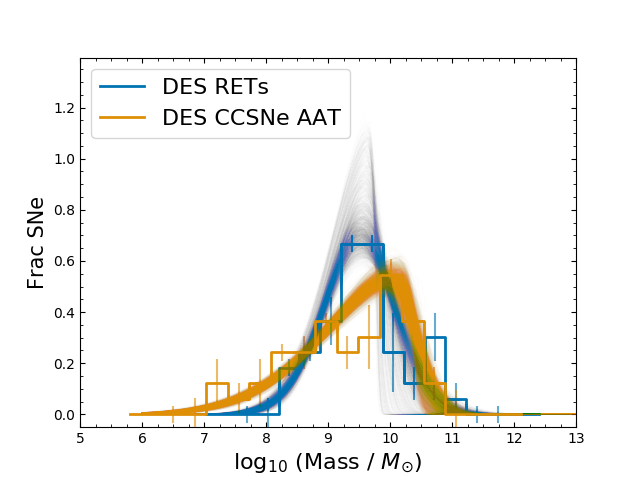}
    \caption{DES RETs compared to DES CCSNe}
    \label{fig:des_RETs_vs_DES_CCSN}
 \end{wrapfigure}
 \vspace{-0.3in}
 \begin{wrapfigure}{1}{1.\linewidth}
    \includegraphics[width=\hsize]{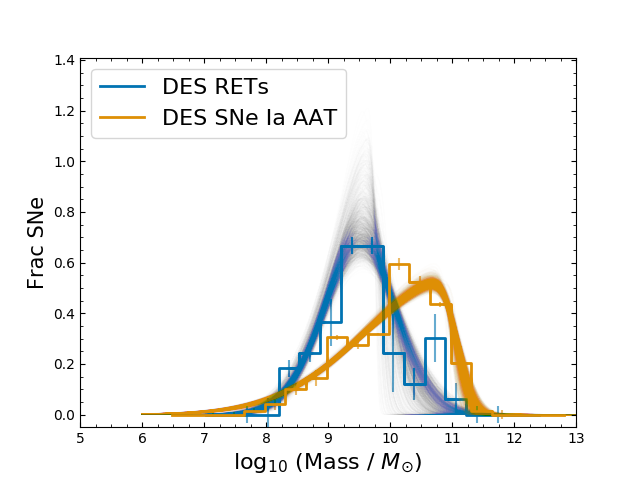}
    \caption{DES RETs compared to DES SNe Ia classified by OzDES }
    \label{fig:des_RETs_vs_DES_SNeIa}
\end{wrapfigure}
\vspace{-0.3in}
\begin{wrapfigure}{1}{1.\linewidth}
    \includegraphics[width=\hsize]{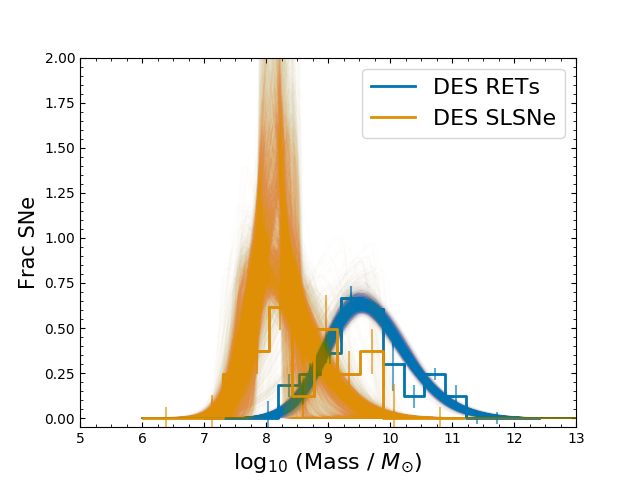}
    \caption{DES RETs compared to DES SLSNe. }
    \label{fig:des_RETs_vs_DES_SLSNe}
 \end{wrapfigure}

\section{Stellar Mass Fit Parameters}\label{app:fitparams}

The following tables show the results from the two-sample Bayesian fits as described in Section \ref{subsec:res_stats}. Loc, scale, and alpha correspond to the location (i.e. central mass), width, and skewness of the distributions respectively. Negative values in the difference parameters mean the DES sample (blue in Appendix C) has a lower value than the comparison sample (yellow in Appendix C). The columns mean, std, and mc\_error correspond to the inferred mean and standard deviation of the posterior distribution, and the simulation standard error from the MCMC, respectively.

\begin{wraptable}{1}{1.\linewidth}
\caption{Results of the MCMC fits for comparing the stellar mass distributions for SNe Ia}

\begin{tabular}{lrrrrr}
\toprule
{} &   mean &    std &  mc\_error \\
\midrule
DES SNe Ia AAT loc   & 11.083 & 0.037 &     0.001 \\
DES SNe Ia AAT scale &  1.462 & 0.036 &     0.001 \\
DES SNe Ia AAT alpha & -8.165 & 1.820 &     0.055 \\
PTF SNe Ia loc       & 11.140 & 0.030 &     0.001 \\
PTF SNe Ia scale     &  1.274 & 0.026 &     0.001 \\
PTF SNe Ia alpha     & -4.694 & 2.336 &     0.127 \\
\midrule
difference of locs   & -0.057 & 0.048 &     0.001 \\
difference of scales &  0.188 & 0.044 &     0.001 \\
difference of alphas & -3.471 & 2.957 &     0.136 \\
\bottomrule
\end{tabular}

\begin{tabular}{lrrrrr}
\toprule
{} &   mean &    std &  mc\_error \\
\midrule
DES SNe Ia AAT loc   & 11.085 & 0.037 &     0.001 \\
DES SNe Ia AAT scale &  1.463 & 0.036 &     0.001 \\
DES SNe Ia AAT alpha & -8.096 & 1.719 &     0.051 \\
PS1 SNe Ia loc       & 11.555 & 0.035 &     0.001 \\
PS1 SNe Ia scale     &  1.804 & 0.052 &     0.001 \\
PS1 SNe Ia alpha     & -5.740 & 0.864 &     0.019 \\
\midrule
difference of locs   & -0.470 & 0.050 &     0.001 \\
difference of scales & -0.341 & 0.063 &     0.001 \\
difference of alphas & -2.356 & 1.923 &     0.055 \\
\bottomrule
\end{tabular}

\begin{tabular}{lrrrrr}
\toprule
{} &   mean &    std &  mc\_error\\
\midrule
DES SNe Ia AAT loc   & 11.082 & 0.038 &     0.001 \\
DES SNe Ia AAT scale &  1.461 & 0.037 &     0.001 \\
DES SNe Ia AAT alpha & -8.253 & 2.009 &     0.077 \\
SNLS SNe Ia loc      & 11.152 & 0.020 &     0.001 \\
SNLS SNe Ia scale    &  1.549 & 0.030 &     0.001 \\
SNLS SNe Ia alpha    & -3.400 & 0.311 &     0.009 \\
\midrule

difference of locs   & -0.070 & 0.043 &     0.001 \\
difference of scales & -0.088 & 0.048 &     0.001 \\
difference of alphas & -4.854 & 2.052 &     0.079 \\
\bottomrule
\end{tabular}
\end{wraptable}

\begin{wraptable}{1}{1.\linewidth}
\caption{Results of the MCMC fits for comparing the stellar mass distrutions for CCSNe}

\begin{tabular}{lrrrrr}
\toprule
{} &    mean &    std &  mc\_error &\\
\midrule
DES CCSNe AAT loc   & 10.440 & 0.063 &     0.001 \\
DES CCSNe AAT scale &  1.371 & 0.066 &     0.001 \\
DES CCSNe AAT alpha & -7.000 & 3.575 &     0.129 \\
SDSS CCSNe loc      & 10.444 & 0.072 &     0.002 \\
SDSS CCSNe scale    &  1.606 & 0.060 &     0.001 \\
SDSS CCSNe alpha    & -4.291 & 2.068 &     0.103 \\
\midrule
difference of locs   & -0.004 & 0.093 &     0.002 \\
difference of scales & -0.235 & 0.089 &     0.002 \\
difference of alphas & -2.709 & 4.150 &     0.166 \\
\bottomrule
\end{tabular}
\end{wraptable}

\begin{wraptable}{1}{1.\linewidth}
\caption{Results of the MCMC fits for comparing the stellar mass distributions for RETs}
\begin{tabular}{lrrrrr}
\toprule
{} &   mean &    std &  mc\_error \\
\midrule
DES RETs loc   &  9.738 & 0.288 &     0.019 \\
DES RETs scale &  0.714 & 0.077 &     0.004 \\
DES RETs alpha & -1.507 & 2.502 &     0.162 \\
PS1 RETs loc   &  8.749 & 0.294 &     0.016 \\
PS1 RETs scale &  0.495 & 0.225 &     0.012 \\
PS1 RETs alpha &  2.369 & 7.841 &     0.480 \\
\midrule

difference of locs   &  0.989 & 0.410 &     0.026 \\
difference of scales &  0.219 & 0.236 &     0.012 \\
difference of alphas & -3.876 & 8.453 &     0.526 \\
\bottomrule
\end{tabular}

\begin{tabular}{lrrrrr}
\toprule
{} &   mean &    std &  mc\_error & \\
\midrule
DES RETs loc        &  9.744 & 0.282 &     0.016 \\
DES RETs scale      &  0.715 & 0.075 &     0.003 \\
DES RETs alpha      & -1.778 & 2.877 &     0.155 \\
DES CCSNe AAT loc   & 10.444 & 0.066 &     0.002 \\
DES CCSNe AAT scale &  1.373 & 0.067 &     0.002 \\
DES CCSNe AAT alpha & -6.566 & 2.967 &     0.119 \\
\midrule
difference of locs   & -0.701 & 0.287 &     0.016 \\
difference of scales & -0.658 & 0.100 &     0.003 \\
difference of alphas &  4.788 & 4.165 &     0.211 \\
\bottomrule
\end{tabular}

\begin{tabular}{lrrrrr}
\toprule
{} &   mean &    std &  mc\_error & \\
\midrule
DES RETs loc         &  9.766 & 0.272 &     0.017 \\
DES RETs scale       &  0.717 & 0.076 &     0.003 \\
DES RETs alpha       & -1.572 & 2.216 &     0.123 \\
DES SNe Ia AAT loc   & 11.084 & 0.037 &     0.001 \\
DES SNe Ia AAT scale &  1.462 & 0.037 &     0.001 \\
DES SNe Ia AAT alpha & -8.135 & 1.869 &     0.063 \\
\midrule

difference of locs   & -1.318 & 0.274 &     0.017 \\
difference of scales & -0.745 & 0.084 &     0.003 \\
difference of alphas &  6.563 & 2.923 &     0.140 \\
\bottomrule
\end{tabular}

\begin{tabular}{lrrrrr}
\toprule
{} &   mean &    std &  mc\_error  \\
\midrule
DES RETs loc    & 9.555 & 1.099 &     0.108 \\
DES RETs scale  & 0.745 & 0.309 &     0.030 \\
DES RETs alpha  & 0.991 & 3.758 &     0.186 \\
DES SLSNe loc   & 7.975 & 0.396 &     0.025 \\
DES SLSNe scale & 0.625 & 0.203 &     0.010 \\
DES SLSNe alpha & 1.094 & 6.614 &     0.356 \\

\midrule
difference of locs   &  1.580 & 1.198 &     0.114 \\
difference of scales &  0.120 & 0.382 &     0.033 \\
difference of alphas & -0.103 & 7.720 &     0.419 \\
\bottomrule
\end{tabular}
\end{wraptable}
\begin{wraptable}{1}{1.\linewidth}
\caption{Results of the MCMC fits for comparing the stellar mass distributions for SLSNe}

\begin{tabular}{lrrrrr}
\toprule
{} &   mean &    std &  mc\_error\\
\midrule
DES SLSNe loc   & 7.987 & 0.365 &     0.020 \\
DES SLSNe scale & 0.597 & 0.208 &     0.010 \\
DES SLSNe alpha & 0.646 & 6.604 &     0.338 \\
PTF SLSNe loc   & 8.309 & 0.447 &     0.020 \\
PTF SLSNe scale & 1.275 & 0.161 &     0.007 \\
PTF SLSNe alpha & 1.444 & 1.759 &     0.078 \\
\midrule

difference of locs   & -0.323 & 0.576 &     0.028 \\
difference of scales & -0.678 & 0.264 &     0.013 \\
difference of alphas & -0.798 & 6.797 &     0.338 \\
\bottomrule
\end{tabular}

\begin{tabular}{lrrrrr}
\toprule
{} &   mean &    std &  mc\_error \\
\midrule
DES SLSNe loc   & 7.897 & 0.331 &     0.020 \\
DES SLSNe scale & 0.644 & 0.208 &     0.011 \\
DES SLSNe alpha & 2.029 & 6.017 &     0.353 \\
PS+ SLSNe loc   & 8.414 & 0.464 &     0.020 \\
PS+ SLSNe scale & 0.913 & 0.107 &     0.003 \\
PS+ SLSNe alpha & 0.040 & 0.997 &     0.042 \\
\midrule
difference of locs   & -0.517 & 0.567 &     0.028 \\
difference of scales & -0.268 & 0.234 &     0.012 \\
difference of alphas &  1.989 & 6.110 &     0.357 \\
\bottomrule
\end{tabular}

\begin{tabular}{lrrrrr}
\toprule
{} &   mean &    std &  mc\_error \\
\midrule
DES SLSNe loc       &  7.987 & 0.421 &     0.024 \\
DES SLSNe scale     &  0.623 & 0.208 &     0.010 \\
DES SLSNe alpha     &  0.773 & 6.462 &     0.329 \\
DES CCSNe AAT loc   & 10.378 & 0.061 &     0.001 \\
DES CCSNe AAT scale &  1.413 & 0.049 &     0.001 \\
DES CCSNe AAT alpha & -6.892 & 3.468 &     0.114 \\
\midrule
difference of locs   & -2.391 & 0.426 &     0.024 \\
difference of scales & -0.790 & 0.212 &     0.010 \\
difference of alphas &  7.665 & 7.466 &     0.359 \\
\bottomrule
\end{tabular}
\end{wraptable}
\end{multicols}

\parbox{\textwidth}{
$^{1}$ School of Physics and Astronomy, University of Southampton,  Southampton, SO17 1BJ, UK\\
$^{2}$ Universit\'e Clermont Auvergne, CNRS/IN2P3, LPC, F-63000 Clermont-Ferrand, France\\
$^{3}$ Lawrence Berkeley National Laboratory, 1 Cyclotron Road, Berkeley, CA 94720, USA\\
$^{4}$ Institute of Cosmology and Gravitation, University of Portsmouth, Portsmouth, PO1 3FX, UK\\
$^{5}$ University of Copenhagen, Dark Cosmology Centre, Juliane Maries Vej 30, 2100 Copenhagen O, Denmark\\
$^{6}$ NASA Einstein Fellow\\
$^{7}$ Department of Physics and Astronomy, University of Pennsylvania, Philadelphia, PA 19104, USA\\
$^{8}$ School of Mathematics and Physics, University of Queensland,  Brisbane, QLD 4072, Australia\\
$^{9}$ Santa Cruz Institute for Particle Physics, Santa Cruz, CA 95064, USA\\
$^{10}$ PITT PACC, Department of Physics and Astronomy, University of Pittsburgh, Pittsburgh, PA 15260, USA\\
$^{11}$ Department of Astronomy and Astrophysics, University of Chicago, Chicago, IL 60637, USA\\
$^{12}$ Kavli Institute for Cosmological Physics, University of Chicago, Chicago, IL 60637, USA\\
$^{13}$ Sydney Institute for Astronomy, School of Physics, A28, The University of Sydney, NSW 2006, Australia\\
$^{14}$ The Research School of Astronomy and Astrophysics, Australian National University, ACT 2601, Australia\\
$^{15}$ Department of Physics, Duke University Durham, NC 27708, USA\\
$^{16}$ Cerro Tololo Inter-American Observatory, National Optical Astronomy Observatory, Casilla 603, La Serena, Chile\\
$^{17}$ Departamento de F\'isica Matem\'atica, Instituto de F\'isica, Universidade de S\~ao Paulo, CP 66318, S\~ao Paulo, SP, 05314-970, Brazil\\
$^{18}$ Laborat\'orio Interinstitucional de e-Astronomia - LIneA, Rua Gal. Jos\'e Cristino 77, Rio de Janeiro, RJ - 20921-400, Brazil\\
$^{19}$ Fermi National Accelerator Laboratory, P. O. Box 500, Batavia, IL 60510, USA\\
$^{20}$ Instituto de Fisica Teorica UAM/CSIC, Universidad Autonoma de Madrid, 28049 Madrid, Spain\\
$^{21}$ CNRS, UMR 7095, Institut d'Astrophysique de Paris, F-75014, Paris, France\\
$^{22}$ Sorbonne Universit\'es, UPMC Univ Paris 06, UMR 7095, Institut d'Astrophysique de Paris, F-75014, Paris, France\\
$^{23}$ Department of Physics \& Astronomy, University College London, Gower Street, London, WC1E 6BT, UK\\
$^{24}$ Kavli Institute for Particle Astrophysics \& Cosmology, P. O. Box 2450, Stanford University, Stanford, CA 94305, USA\\
$^{25}$ SLAC National Accelerator Laboratory, Menlo Park, CA 94025, USA\\
$^{26}$ Centro de Investigaciones Energ\'eticas, Medioambientales y Tecnol\'ogicas (CIEMAT), Madrid, Spain\\
$^{27}$ Department of Astronomy, University of Illinois at Urbana-Champaign, 1002 W. Green Street, Urbana, IL 61801, USA\\
$^{28}$ National Center for Supercomputing Applications, 1205 West Clark St., Urbana, IL 61801, USA\\
$^{29}$ Observat\'orio Nacional, Rua Gal. Jos\'e Cristino 77, Rio de Janeiro, RJ - 20921-400, Brazil\\
$^{30}$ Department of Physics, IIT Hyderabad, Kandi, Telangana 502285, India\\
$^{31}$ Department of Astronomy/Steward Observatory, University of Arizona, 933 North Cherry Avenue, Tucson, AZ 85721-0065, USA\\
$^{32}$ Jet Propulsion Laboratory, California Institute of Technology, 4800 Oak Grove Dr., Pasadena, CA 91109, USA\\
$^{33}$ Institut d'Estudis Espacials de Catalunya (IEEC), 08034 Barcelona, Spain\\
$^{34}$ Institute of Space Sciences (ICE, CSIC),  Campus UAB, Carrer de Can Magrans, s/n,  08193 Barcelona, Spain\\
$^{35}$ Department of Astronomy, University of Michigan, Ann Arbor, MI 48109, USA\\
$^{36}$ Department of Physics, University of Michigan, Ann Arbor, MI 48109, USA\\
$^{37}$ Department of Physics, ETH Zurich, Wolfgang-Pauli-Strasse 16, CH-8093 Zurich, Switzerland\\
$^{38}$ Center for Astrophysics $\vert$ Harvard \& Smithsonian, 60 Garden Street, Cambridge, MA 02138, USA\\
$^{39}$ Australian Astronomical Optics, Macquarie University, North Ryde, NSW 2113, Australia\\
$^{40}$ Lowell Observatory, 1400 Mars Hill Rd, Flagstaff, AZ 86001, USA\\
$^{41}$ Center for Cosmology and Astro-Particle Physics, The Ohio State University, Columbus, OH 43210, USA\\
$^{42}$ Department of Astronomy, The Ohio State University, Columbus, OH 43210, USA\\
$^{43}$ Department of Astrophysical Sciences, Princeton University, Peyton Hall, Princeton, NJ 08544, USA\\
$^{44}$ Instituci\'o Catalana de Recerca i Estudis Avan\c{c}ats, E-08010 Barcelona, Spain\\
$^{45}$ Institut de F\'{\i}sica d'Altes Energies (IFAE), The Barcelona Institute of Science and Technology, Campus UAB, 08193 Bellaterra (Barcelona) Spain\\
$^{46}$ Department of Physics and Astronomy, Pevensey Building, University of Sussex, Brighton, BN1 9QH, UK\\
$^{47}$ Computer Science and Mathematics Division, Oak Ridge National Laboratory, Oak Ridge, TN 37831\\
$^{48}$ Max Planck Institute for Extraterrestrial Physics, Giessenbachstrasse, 85748 Garching, Germany\\
$^{49}$ Universit\"ats-Sternwarte, Fakult\"at f\"ur Physik, Ludwig-Maximilians Universit\"at M\"unchen, Scheinerstr. 1, 81679 M\"unchen, Germany\\
}





\bsp	
\label{lastpage}
\end{document}